\documentclass[%
 reprint,
superscriptaddress,
bibnotes,
 amsmath,amssymb,
 aps,
prb,
]{revtex4-1}

\usepackage{graphicx}
\usepackage{dcolumn}
\usepackage{bm}
\usepackage{color}
\definecolor{OliveGreen}{rgb}{0,0.6,0}
\definecolor{auburn}{rgb}{0.43, 0.21, 0.1}
\definecolor{blue_violet}{rgb}{0.54, 0.17, 0.89}

\usepackage{appendix}
\usepackage{chngcntr}
\usepackage{etoolbox}
\usepackage{lipsum}

\AtBeginEnvironment{appendices}{%
\addcontentsline{toc}{section}{Appendices}
\counterwithin{figure}{section}
\counterwithin{equation}{section}
\counterwithin{table}{section}
}

\begin{document}


\title{Kinetic Pathways of Phase Decomposition Using Steepest-Entropy-Ascent Quantum Thermodynamics Modeling. Part\;I: Continuous and Discontinuous Transformations}

\author{Ryo Yamada}
\email{ryo213@vt.edu}
\affiliation{Materials Science and Engineering Department, Virginia Polytechnic Institute and State University, Blacksburg, Virginia 24061, USA}
\author{Michael R. von Spakovsky}
\email{vonspako@vt.edu}
\affiliation{Center for Energy Systems Research, Mechanical Engineering Department, Virginia Polytechnic Institute and State University, Blacksburg, Virginia 24061, USA}
\author{William T. Reynolds, Jr.}
\email{reynolds@vt.edu}
\affiliation{Materials Science and Engineering Department, Virginia Polytechnic Institute and State University, Blacksburg, Virginia 24061, USA}

\date{\today}

\begin{abstract}
The decomposition kinetics of a solid-solution into separate phases are analyzed with an equation of motion initially developed to account for dissipative processes in quantum systems. This equation and the steepest-entropy-ascent quantum thermodynamic framework of which it is a part make it possible to track kinetic processes in systems in non-equilibrium, while retaining the framework of classical equilibrium thermodynamics. The general equation of motion is particularized for the case of the decomposition of a binary alloy, and a solution model is used to build an approximate energy eigenstructure, or pseudo-eigenstructure, for the alloy system. This equation is then solved with the pseudo-eigenstructure to obtain a unique reaction path and the decomposition kinetics of the alloy. For a hypothetical solid-solution with a miscibility gap at low temperatures, the conditions under which this framework predicts a continuous transformation path (spinodal decomposition) or a discontinuous one (nucleation and growth) are demonstrated.
\end{abstract}

\pacs{Valid PACS appear here}
\maketitle

\section{\label{chap5_sec:level1}Introduction}
J.\;W.\;Gibbs envisioned uniform solutions decomposing (or phase separating) through two kinds of kinetic processes \cite{gibbs1906scientific,cahn1961spinodal}. In alloy systems, these processes are sometimes classified as continuous and discontinuous transformations. While continuous transformations begin with small fluctuations that extend over relatively large spatial regions and take place simultaneously throughout the volume of the system, discontinuous transformations initiate with localized concentration fluctuations that are comparatively large in amplitude but small in spatial extent \cite{balluffi2005kinetics}. From the perspective of the thermodynamic free-energy \cite{cahn1961spinodal,balluffi2005kinetics}, continuous transformations initiate spontaneously from an unstable solution when an infinitesimal variation decreases the free-energy. This behavior is associated with the spinodal decomposition mechanism. Discontinuous transformations develop in an initially metastable solution through a series of statistical fluctuations that eventually overcome a free-energy barrier. They are characteristic of nucleation and growth mechanisms. These thermodynamic concepts are useful for interpreting alloy decomposition even though functions like temperature and free-energy are strictly speaking defined only at equilibrium and must be extrapolated to non-equilibrium states to describe kinetic phenomena.

Although the unit process underlying the two mechanisms are the same (atomic migration by diffusion), the driving forces are quite different, and this leads to very different kinetic characteristics. Models for decomposition processes like these generally start by assuming that a particular step of the process is rate-limiting, and then building an appropriate mathematical description of the rate-limiting step. An inherent difficulty with this approach is the need to know the underlying reaction mechanism in order to build an accurate kinetic model. For example, if classical nucleation is the operative process responsible for phase decomposition, the kinetics are described in terms of the distribution of cluster sizes and their rates of growth and shrinkage \cite{schmelzer2004dynamics}. On the other hand, if spinodal decomposition is operative, the decomposition rate is better described by a generalized diffusion equation (e.g., reference \cite{cahn1961spinodal}). For this reason, microstructural modeling starts by assuming a decomposition mechanism rather than determining it from the physical conditions. 

Following Gibbs \cite{gibbs1906scientific}, the decomposition mechanism should be selected at the very beginning of the decomposition process when changes take place through the collective behavior of a relatively small number of fluctuations. Not surprisingly, kinetic Monte Carlo methods, which are based on statistical fluctuations and do not assume a rate-limiting step, are successful describing multiple processes \cite{soisson2006kinetic,gao2018theoretical}. Quantum mechanics is widely used to interpret discrete behavior in small systems, so it should be reasonable to apply  the tools of quantum mechanics to the {\it selection} of transformation mechanisms in bulk systems. 

In this regard, the steepest-entropy-ascent quantum thermodynamics (SEAQT) framework shows great promise for predicting both the operative decomposition mechanism as well as the reaction kinetics. SEAQT is a non-equilibrium thermodynamic-ensemble approach that was originally formulated to address a number of physical inconsistencies between quantum mechanics and thermodynamics \cite{hatsopoulos1976-I,hatsopoulos1976-IIa,hatsopoulos1976-IIb,hatsopoulos1976-III,beretta2005generalPhD}. It describes the relaxation process of a system from an initial non-equilibrium state to stable equilibrium following the direction of steepest entropy ascent, i.e., maximum entropy production. To apply the framework to the phase decomposition of alloys, the system is described differently from conventional microstructural models. Rather than describing the system in terms of position-dependent functions, like free-energy, that evolve with time, the SEAQT approach employs a thermodynamic-ensemble and a density operator formalism (analogous to a phase-space probability measure in statistical mechanics) that tracks the decomposition process in terms of a single time-dependent variable. While perhaps physically nonintuitive, reformulating the problem in this way has important computational advantages over approaches based on classical mechanics (e.g., molecular dynamics) and microstructual models (e.g., phase field models). 

\begin{figure}
\begin{center}
\includegraphics[scale=0.34]{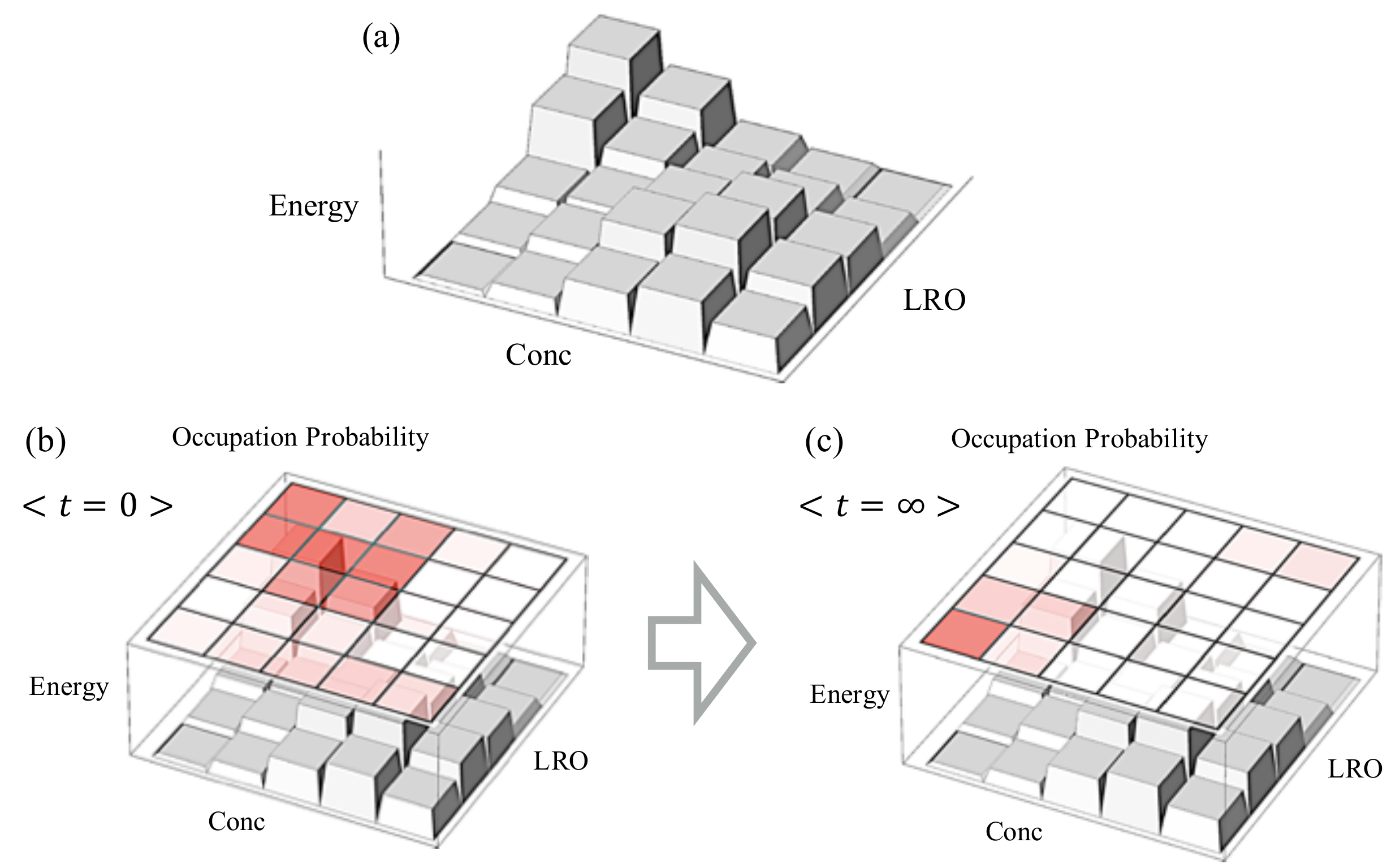}
\caption{\label{fig5:SEAQT_flowchart} A schematic explanation of the SEAQT approach: (a) An energy landscape, or eigenstructure, of an alloy with variable composition and long-range order is constructed from an appropriate solution model. The energy of the system is displayed as a discrete function of alloy concentration and a long-range order parameter (LRO). (b) The initial state of the system ($t=0$) is expressed by occupation probabilities for each possible configuration, which is superimposed over the eigenstructure (shaded squares). The time-evolution of the system is determined by solving the SEAQT equation of motion (represented by the large arrow) to find the path from the initial state to that of stable equilibrium (c) at $t=\infty$.}
\end{center}
\end{figure}

States in the SEAQT framework are described by occupation probabilities of a set of possible energy eigenlevels, also called the energy eigenstructure \cite{li2016steepest}, as depicted in Fig.\;\ref{fig5:SEAQT_flowchart}. For example, an energy eigenstructure for a A--B binary solid-solution of a specified size is constructed from the energies corresponding to all the possible arrangements of A-type and B-type atoms. The entropy of the system is given by a measure of the degree of energy load sharing among available energy eigenlevels, and the evolution of the system from an initial, non-equilibrium state at time $t=0$ to a final, stable equilibrium state at time $t=\infty$ is found by solving the SEAQT equation of motion (indicated by the large schematic arrow in Fig.\;\ref{fig5:SEAQT_flowchart}).  By assuming the system's evolution of state follows the path of steepest entropy ascent (maximum rate of entropy production), the equation of motion yields a unique kinetic path through state space from the initial state to the final equilibrium state predicted by the second law of thermodynamics.

To use the SEAQT framework, the energy eigenstructure must be determined for the system in question. Although the eigenstructure for a gas phase can be constructed relatively easily (e.g., by assuming ideal gas behavior), many-body interactions among particles make the eigenstructure highly complex for condensed phases. There are two aspects to this complexity. First, determining the available energy eigenlevels from appropriate quantum models may be computationally intractable, and second, the number of energy eigenlevels is effectively infinite. Both of these problems are addressed in recent work modeling the thermal expansion of silver \cite{yamada2018method}. A highly simplified eigenstructure is built from a reduced-order model  (a solid-state instead of a quantum model), and an infinite energy-level eigenstructure is replaced with a discretized, finite-level ``pseudo-eigenstructure'' with the use of the density of states method developed in reference \cite{li2016steepest}.

In this contribution, the SEAQT theoretical framework with the pseudo-eigenstructure is applied to phase decompositions in binary solid-solutions to determine the kinetic pathways. The work consists of two parts; continuous and discontinuous transformations are investigated in Part\;I, and ordering and concurrent ordering with phase separation are explored in Part\;II. Part\;I is organized as follows. First, the SEAQT equation of motion is modified for kinetic calculations in binary alloy systems with fixed composition in Sec.\;\ref{chap5_sec:level2_1}, and a pseudo-eigenstructure for a solid-solution is constructed using a mean-field approximation (or a solution model) in Sec.\;\ref{chap5_sec:level2_2}. In Sec.\;\ref{chap5_sec:level2_3}, calculation conditions and how to prepare initial states are described. In Sec.\;\ref{chap5_sec:level3}, the calculated time-evolution of the decomposition process from arbitrary initial states is shown and discussed focusing on the continuous and discontinuous transformation behaviors (Sec.\;\ref{chap5_sec:level3_1}) in which a spinodal limit and a real time-dependence of the decomposition process are also explored (Secs.\;\ref{chap5_sec:level3_2} and \ref{chap5_sec:level3_3}, respectively). At the end, the study of the continuous and discontinuous phase decomposition behaviors in an alloy system using the SEAQT model is summarized in Sec.\;\ref{chap5_sec:level4}.

\section{\label{chap5_sec:level2}Theory}

\subsection{\label{chap5_sec:level2_1}SEAQT equation of motion}
The equation of motion in the SEAQT modeling has been developed to account for dissipative processes in quantum systems. The dissipative contribution is incorporated in the Schr\"{o}dinger equation as the irreversible term and the SEAQT equation of motion takes a form \cite{beretta1985quantum,beretta2006nonlinear,beretta2009nonlinear}:
\begin{equation}
\frac{d\hat{\rho}}{dt}=\frac{1}{i\hbar}[\hat{\rho},\hat{H}]+\frac{1}{\tau(\hat{\rho})}\hat{D}(\hat{\rho}) \; , \label{eq5:equation_of_motion}
\end{equation}
where $\hat{\rho}$ is the density operator, $t$ the time, $\hbar$ the reduced Planck constant, $\hat{H}$ the Hamiltonian operator, $\tau$ the relaxation time (which represents the rate at which the states of a system evolve in Hilbert space along the unique kinetic path determined by Eq.\;(\ref{eq5:equation_of_motion})), and $\hat{D}$ the dissipation operator. The left-hand side of the equation and the first term on the right corresponds to the time-dependent von Neumann (or Schr\"{o}dinger) equation. The second term on the right is a dissipation term, the irreversible contribution that accounts for relaxation processes in the system. When $\hat{\rho}$ is diagonal in the Hamiltonian eigenvector basis, $\hat{\rho}$ and $\hat{H}$ commute and the von Neumann term in the equation of motion disappears so that Eq.\;(\ref{eq5:equation_of_motion}) simplifies (for the case of a system in which the identity and Hamiltonian operators are the only generators of the motion) to \cite{beretta2006nonlinear,beretta2009nonlinear,li2016steepest}
\begin{equation}
\frac{dp_j}{dt}=\frac{1}{\tau}\frac{\begin{vmatrix} 
-p_j \mathrm{ln} \frac{p_j}{g_j} & p_j & \epsilon_jp_j \\
\langle s \rangle & 1 & \langle e \rangle \\
\langle es \rangle & \langle e \rangle & \langle e^2 \rangle
\end{vmatrix}}{\begin{vmatrix} 
1 & \langle e \rangle \\
\langle e \rangle & \langle e^2 \rangle 
\end{vmatrix}} \; ,  \label{eq5:equation_of_motion_simplified}
\end{equation}
where 
\[
\begin{array}{c c}
\langle s \rangle = - \sum\limits_{i} p_i \mathrm{ln} \frac{p_i}{g_i}  \; ,
&
\langle e \rangle = \sum\limits_{i} \epsilon_i p_i  \; , \\ \\
\langle e^2 \rangle = \sum\limits_{i} \epsilon_i^2 p_i \; , \;\;\;
&
\langle es \rangle = - \sum\limits_{i} \epsilon_i p_i \mathrm{ln} \frac{p_i}{g_i}  \; ,
\end{array}
\]
and the $p_j$ are the diagonal terms of $\hat{\rho}$, each of which represents the occupation probability in the $j^{th}$ energy eigenlevel ${\epsilon}_j$; the $g_j$ are the degeneracies of the energy eigenlevel; and $\langle \cdot \rangle$ is the expectation value of the property. Note that the von Neumann formula for entropy is used here. Provided the density operator is based on a homogeneous ensemble, this formula satisfies all the characteristics of entropy required by thermodynamics without making entropy a statistical property of the ensemble \cite{gyftopoulos1997entropy,cubukcu1993thermodynamics,yamada2018steepest}. It is assumed here that $\hat{\rho}$ is diagonal in the eigenvector basis, which is the case for many classical systems or when no quantum correlations between particles are present \cite{li2016generalized,li2016modeling,li2017study}.

The SEAQT equation of motion, Eq.\;(\ref{eq5:equation_of_motion_simplified}), is derived via a constrained gradient in Hilbert space that causes the system to follow the direction of steepest entropy ascent when the energy and occupation probabilities are conserved. When the number of particles is conserved as an additional constraint, the identity, Hamiltonian, and particle number operators become the generators of the motion. The equation of motion, then, becomes \cite{li2016steepest2}
\begin{equation}
\frac{dp_j}{dt}=\frac{1}{\tau}\frac{\begin{vmatrix} 
-p_j \mathrm{ln} \frac{p_j}{g_j} & p_j &  N_j p_j & \epsilon_jp_j \\
\langle s \rangle & 1 & \langle N \rangle & \langle e \rangle \\
\langle Ns \rangle & \langle N \rangle & \langle N^2 \rangle & \langle eN \rangle \\
\langle es \rangle & \langle e \rangle & \langle eN \rangle & \langle e^2 \rangle
\end{vmatrix}}{\begin{vmatrix} 
 1 & \langle N \rangle & \langle e \rangle \\
 \langle N \rangle & \langle N^2 \rangle & \langle eN \rangle \\
 \langle e \rangle & \langle eN \rangle & \langle e^2 \rangle
\end{vmatrix}} \;  ,  \label{eq5:equation_of_motion_grand_canonical}
\end{equation}
where 
\[
\begin{array}{c c}
\langle N \rangle = \sum\limits_{i} N_i p_i  \; ,
&
\langle N^2 \rangle = \sum\limits_{i} N_i^2 p_i \; , \\ \\
\langle eN \rangle = \sum\limits_{i} \epsilon_i N_i p_i \; , \;\;\;
&
\langle Ns \rangle = - \sum\limits_{i} N_i p_i \mathrm{ln} \frac{p_i}{g_i}  \; .
\end{array}
\]
Here the $N_j$ are the number of particles in the $j^{th}$ energy eigenlevel. The equation of motion can be modified further by allowing an exchange of heat between the system and a heat reservoir. This can be done by viewing them as subsystems of an overall composite system (see references \cite{li2016steepest,li2016generalized,li2016steepest2,yamada2018steepest}) for which the generators of the motion are the identity and particle number operators for each subsystem and the Hamiltonian operator for the composite system. This combined with the concept of hypoequilibrium states \cite{li2016steepest,li2016generalized,li2016steepest2} transforms Eq.\;(\ref{eq5:equation_of_motion_grand_canonical}) for the original system into the following form:
\begin{equation}
\frac{dp_j}{dt}=\frac{1}{\tau} p_j \left[ \left( s_j - \langle s \rangle \right) + \left( N_j- \langle N \rangle \right) \gamma^R - \left( \epsilon_j - \left< e \right> \right) \beta^R \right] ,     
\label{eq5:equation_motion_grand_canonical_heat}
\end{equation}
where 
\[
\gamma^R \equiv -\frac{ (\langle Ns \rangle - \langle N \rangle \langle s \rangle) - ( \langle eN \rangle - \langle e \rangle \langle N \rangle) \beta^R}{ \langle N^2 \rangle - \langle N \rangle \langle N \rangle} \; ,
\]
and $\beta^R$ is the inverse of the product of Boltzmann's constant and the temperature of the reservoir $T_R$, i.e., $\beta^R=1/k_BT_R$. 

For many physical processes occurring in an alloy, the concentrations of the components remain constant. This can be described for a binary A--B alloy by replacing $N_j$ with $N_{B,j}$ (or $N_{A,j}$) and fixing the total number of particles in each energy eigenlevel (i.e., $N_j=N_{A,j}+N_{B,j}=\mbox{constant}$ where $N_{A,j}$ and $N_{B,j}$ are, respectively, the number of A-type and B-type atoms in the $j^{th}$ energy eigenlevel). These notations together with Eqs.\;(\ref{eq5:equation_of_motion_grand_canonical}) and (\ref{eq5:equation_motion_grand_canonical_heat}) are applicable to a binary alloy of fixed composition. 


\subsection{\label{chap5_sec:level2_2}Pseudo-eigenstructure}
Configurational energy in a binary alloy system is given by \cite{khachaturyan2013theory}
\begin{equation}
E=\frac{1}{2} \sum_{\bm{\mathrm{r}},\bm{\mathrm{r}}'} W(\bm{\mathrm{r}} - \bm{\mathrm{r}}') n(\bm{\mathrm{r}}) n(\bm{\mathrm{r}}') \; ,   \label{eq5:total_energy_MF_original}
\end{equation}
where $W(\bm{\mathrm{r}} - \bm{\mathrm{r}}')$ is a pairwise interatomic interaction energy between two atoms at lattice sites $\bm{\mathrm{r}}$ and $\bm{\mathrm{r}}'$. The factors $n(\bm{\mathrm{r}})$ and $n(\bm{\mathrm{r}}')$ represent the distribution functions at these lattice points. The pseudo-eigenstructure in an alloy system is constructed by employing a mean-field approximation that replaces many-body interactions among particles with an average internal field experienced by each atom \cite{girifalco2003statistical}. Using the simplest mean-field approximation, where short-range correlations between different atomic species are ignored, the $n(\bm{\mathrm{r}})$ and $n(\bm{\mathrm{r}}')$ can be expressed in terms of the concentration of B-type atoms, $c$. When the reference energy is set to the segregation limit (a line connecting the energies of two systems  composed of pure A-type and pure B-type atoms), Eq.\;(\ref{eq5:total_energy_MF_original}) becomes
\begin{equation}
E(c)= \frac{1}{2} N c (1-c) V(\bm{0}) \; ,   \label{eq5:total_energy_phase_separation}  
\end{equation}
where $N$ is the number of atoms in the system and $V(\bm{0})$ is a parameter incorporating all the interaction energies. For a face-centered cubic crystal, $V(\bm{0})$ is given by \cite{khachaturyan2013theory}
\begin{equation}
V(\bm{0})=12w_1+6w_2+24w_3+12w_4+\cdots \;\; ,
\label{eq5:interaction_energies_fcc}
\end{equation}
where $w_{n}$ is the $n^{th}$ nearest-neighbor {\it effective} pair interaction energy, which is related to the component-specific $n^{th}$-neighbor pair interaction energies, $V_{ij}^{(n)}$ ($i,j=$ A or B), by 
\begin{equation}
w_{n}=V_{AA}^{(n)}+V_{BB}^{(n)}-2V_{AB}^{(n)} \;\; .
\label{eq5:effective_interaction_energies_fcc}
\end{equation}
The parameter $V(\bm{0})$ is positive when the interactions among A and B species are such that a solid-solution of A and B prefers to decompose into two different solid-solutions. The degeneracy of each energy in Eq.\;(\ref{eq5:total_energy_phase_separation}) is given by the binomial coefficient,
\begin{equation}
g(c)=\frac{N !}{N_{A} ! \cdot N_{B} !}=\frac{N !}{(N(1-c))! \cdot (Nc)!}  \; ,     \label{eq5:degeneracy_mean_field_binary_alloy}
\end{equation}
where $N_{A}$ and $N_{B}$ are the number of A-type and B-type atoms, respectively. Here, using the approximation for a factorial \cite{weisstein2008stirling}, $x!\approx(2x+\frac{1}{3} \pi) x^x e^{-x}$, Eq.\;(\ref{eq5:degeneracy_mean_field_binary_alloy}) can be treated as a continuous function for large $N$. The energy eigenlevels, $E_j$, and the degeneracy, $g_j$, are determined from Eqs.\;(\ref{eq5:total_energy_phase_separation}) and (\ref{eq5:degeneracy_mean_field_binary_alloy}) by replacing $N$ and $c$ with $N_j$ and $c_j$ (here the energy eigenlevels are denoted by $E_j$ instead of $\epsilon_j$ because the $E_j$'s are extensive quantities). Since the $N_j$ are the same for all energy eigenlevels (because of the constraint mentioned at the end of Sec.\;\ref{chap5_sec:level2_1}), it is denoted as $N$ hereafter. For a bulk sample composed of a vast number of particles, any value of $c_j$ between zero and unity is possible and the number of states is effectively infinite. To cope with this intractable number of accessible energy eigenlevels, the density of states method developed by Li and von Spakovsky within the SEAQT framework \cite{li2016steepest} is used, where similar energy eigenlevels are combined into discrete bins and the computational burden is reduced substantially without affecting the accuracy of the result. With this method, the energy eigenlevels, degeneracies, and concentration of B-type atoms become
\begin{equation}
E_j = \frac{1}{g_j} \int_{\bar{c}_j}^{\bar{c}_{j+1}}g(c) E(c) \; dc   \;,  \label{eq5:energy_eigenvalue_pseudo}
\end{equation}
\begin{equation}
g_j=\int_{\bar{c}_j}^{\bar{c}_{j+1}} g(c) \; dc \;,   \label{eq5:degeneracy_pseudo}
\end{equation}
and
\begin{equation}
c_j = \frac{1}{g_j} \int_{\bar{c}_j}^{\bar{c}_{j+1}} g(c) c \; dc \;,    \label{eq5:fraction_down_spin_pseudo}
\end{equation}
where $\bar{c}_j$ is specified by the number of intervals, $R$, as $\bar{c}_j= j/R$ with $j$ an integer ($j=0, 1, 2, ... \; R$). The number of intervals, $R$, is determined by ensuring the following condition is satisfied \cite{yamada2018steepest}:
\begin{equation}
\frac{1}{\beta} \gg \frac{| E_{j+1}-E_j | }{N}  \; .   \label{eq5:quasi_continuous_condition}
\end{equation}

The size of the system, specified via the number of atoms, $N$, establishes the energy and the degeneracy through Eqs.\;(\ref{eq5:total_energy_phase_separation}) and (\ref{eq5:degeneracy_mean_field_binary_alloy}), respectively. In order to capture quantum effects, the system size should not be so large that it behaves classically but large enough to include important interactions among the constituent atoms --- say 5 to 20 times the interatomic distance for a metallic solid-solution. For most of the subsequent calculations, $N=10^4$ was chosen for the system size although a more detailed analysis than that conducted here could be carried out to determine the most appropriate system size, but that is beyond the present scope.

The system being considered here is analogous to what Gibbs called a ``homogenous part of the given mass'' in his seminal paper on the equilibrium of heterogeneous substances \cite{gibbs1906scientific}.  His homogeneous part is spatially uniform in chemical composition and physical state, and it is a subsystem of the larger isolated system he considers at equilbrium. While a uniform system may seem at odds with the concept of fluctuations, it is entirely consistent with the way a system is represented in the SEAQT framework. Fluctuations, or changes in composition or physical state, in the SEAQT system are reflected by multimode probability distributions among the energy eigenlevels, not by spatial variations in a property.  Gibbs demonstrated that equilibrium is reached when the intensive property values (temperature, pressure, and chemical potential) of each homogeneous part are identical. The SEAQT framework is used here to identify the path by which a part reaches this equilibrium.

\subsection{\label{chap5_sec:level2_3}Specification of initial states}
\begin{figure}
\begin{center}
\includegraphics[scale=0.26]{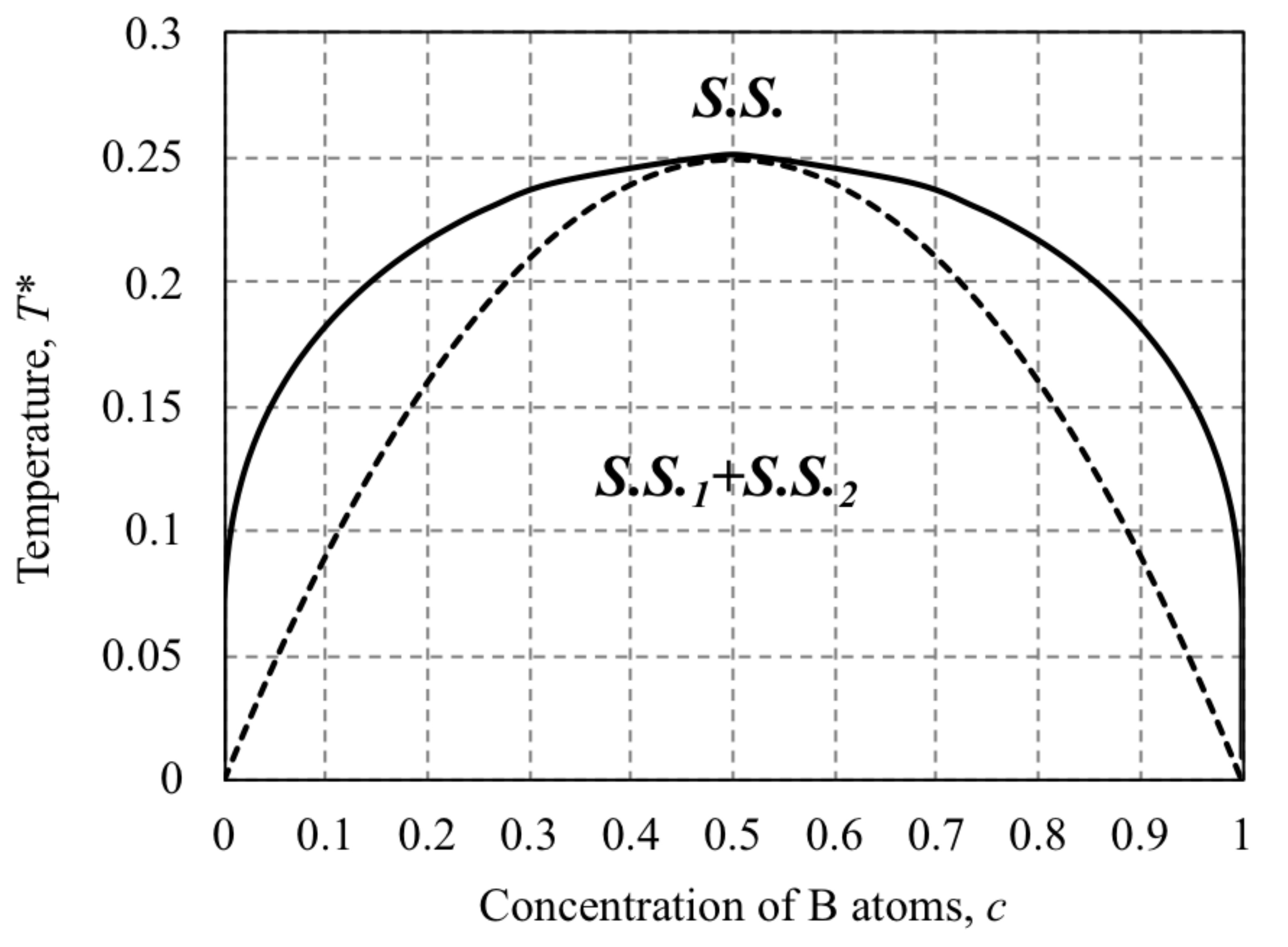}
\caption
{\label{fig5:phase_diagram_phase_separation} A phase diagram with a positive $V(\bm{0})$. The solid line is the solvus line, inside of which is a two-phase region of different solid-solutions. The spinodal curve determined from the free-energy \cite{khachaturyan2013theory} is shown as the broken line. The vertical axis is a normalized temperature, $T^*=\frac{k_BT}{V(\bm{0})}$. } 
\end{center}
\end{figure}

The evolution of a binary solid-solution that is quenched and annealed within a miscibility gap is considered in this work.  The phase diagram for a binary alloy with a high-temperature solid-solution and a miscibility gap at lower temperatures is shown in Fig.\;\ref{fig5:phase_diagram_phase_separation}. The pseudo-eigenstructure of such an alloy corresponds to a system with a positive $V(\bm{0})$ in Eq.\;(\ref{eq5:total_energy_phase_separation}). 

The initial disordered solid-solution ({\it S.S.}) is annealed at a high temperature, $T^H$\;($=T_0$), and then quenched to a lower temperature, $T^L$\;($=T_R$), and annealed at that temperature. The initial state can be prepared using the (semi-) \cite{lesar2013introduction} grand canonical distribution:
\begin{equation}
p^{0}_j=\frac{g_j e^{-\beta^0 ( E_j +\mu_A N_{A,j} +\mu_B N_{B,j} ) }}{\Xi} \; ,  \label{eq5:grand_canonical_distribution}
\end{equation}
where $\beta^0=1/k_BT_0$, $\mu_A$ and $\mu_B$ are, respectively, the chemical potentials of A atoms and B atoms, and $\Xi$ is the grand partition function, which is given by
\begin{equation}
\Xi \equiv \sum\limits_i g_i e^{-\beta^0 ( E_i +\mu_A N_{A,i} +\mu_B N_{B,i} )}  \; . \label{eq5:grand_partition_function}
\end{equation}
The target alloy composition is obtained by adjusting the chemical potentials. Note that one needs to adjust the chemical potentials just for the initial state since once the initial state is prepared using Eq.\;(\ref{eq5:grand_canonical_distribution}), the alloy composition is fixed and conserved in the kinetic calculations via Eq.\;(\ref{eq5:equation_motion_grand_canonical_heat}).  

Although not a necessary assumption, preparing the initial state of the alloy system with Eq.\;(\ref{eq5:grand_canonical_distribution}) alone means that its initial state is in equilibrium at the initial, high temperature, $T_0$. On the other hand, the initial state of the composite system (i.e., alloy system plus reservoir) is that of non-equilibrium since the equilibrium state of the reservoir is not that of the alloy system. This non-equilibrium state is in effect what Li and von Spakovsky \cite{li2016steepest,li2016steepest2} call a $2^{nd}$-order hypoequilibrium state. The concept of hypoequilibrium provides a simple relaxation pattern for a system by properly dividing the system into a number of subsystems (or subspaces). The steepest entropy ascent principle under hypoequilibrium ensures that each subsystem moves along its own manifold of different equilibrium states 
until the states of both subsystems (alloy system plus reservoir) arrive at a final equilibrium state of the composite system in which the two subsystems are in mutual stable equilibrium with each other. In order to explore the effects on state evolution of not assuming that the alloy subsystem is initially in equilibrium, concentration fluctuations are introduced into the initial state to drive it away from equilibrium.
This is done by using an occupation probability distribution corresponding to a smaller number of particles than are actually present in the system, $N_0 < N$. A smaller number of particles reduces the degeneracies of some of the energy eigenlevels, $g_j$, and generates an initial occupation probability distribution calculated from Eq.\;(\ref{eq5:grand_canonical_distribution}) that is broader than the equilibrium distribution. The effects of the number of particles on initial states and kinetic paths are discussed in Sec.\;\ref{chap5_sec:level3_2}.

\section{\label{chap5_sec:level3}Results and Discussion}
\subsection{\label{chap5_sec:level3_1}Continuous and discontinuous transformations}
The SEAQT equation of motion, Eq.\;(\ref{eq5:equation_motion_grand_canonical_heat}), is solved with Eqs.\;(\ref{eq5:energy_eigenvalue_pseudo})\;--\;(\ref{eq5:fraction_down_spin_pseudo}) to track the decomposition process in two alloys, A--40.0\;at.\%\;B and A--30.0\;at.\%\;B, quenched from $T^*_0=\frac{k_BT_0}{V(0)}=0.30$ to $T^*_0=0.20$.  Solving the equation of motion gives the occupancy probabilities of the atomic configurations (distinguished by the concentration of B-type atoms, $c$) as a function of time from the initial state to the final stable equilibrium state.  

\begin{figure}
\includegraphics[scale=0.42]{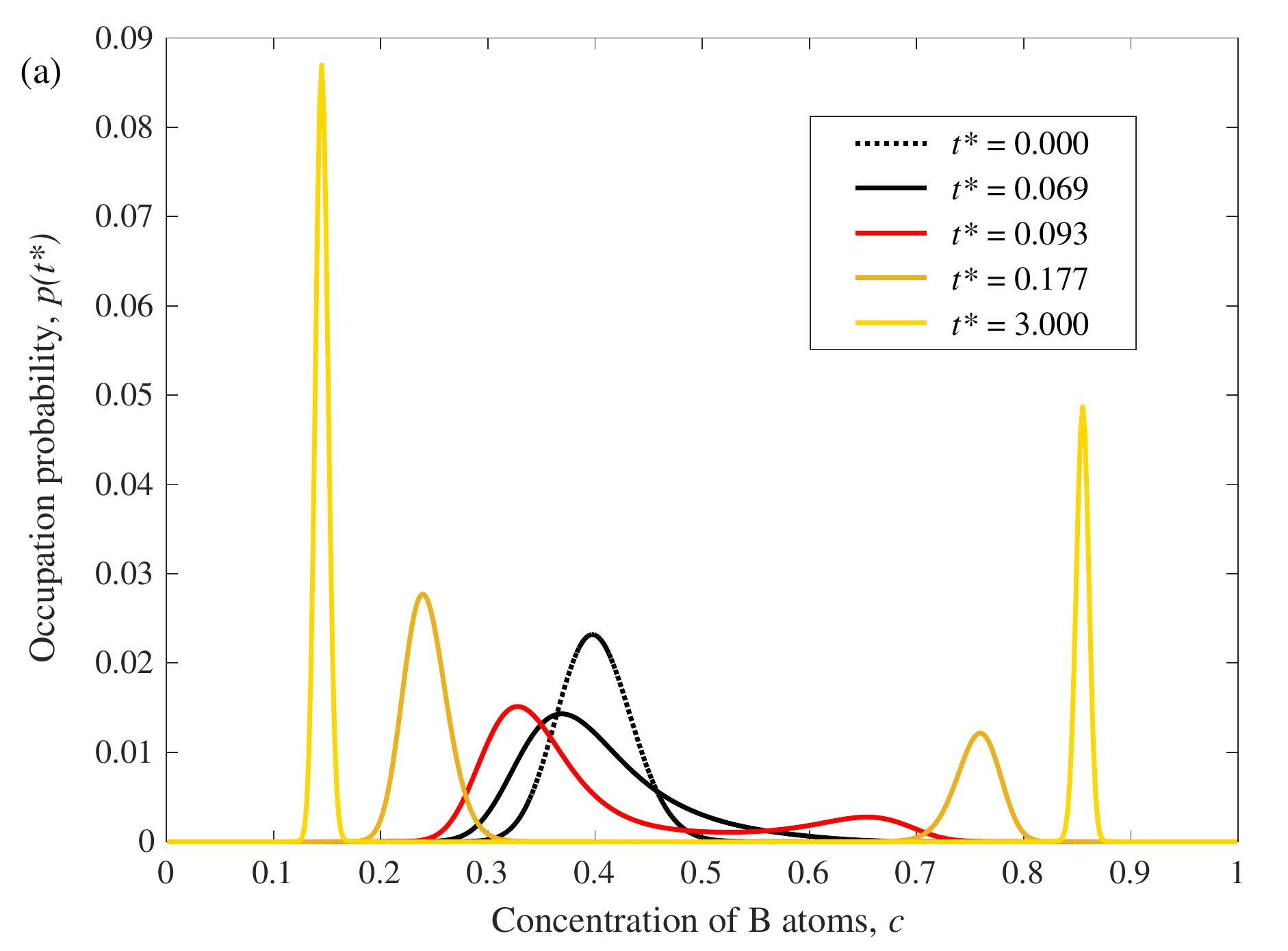}
\includegraphics[scale=0.42]{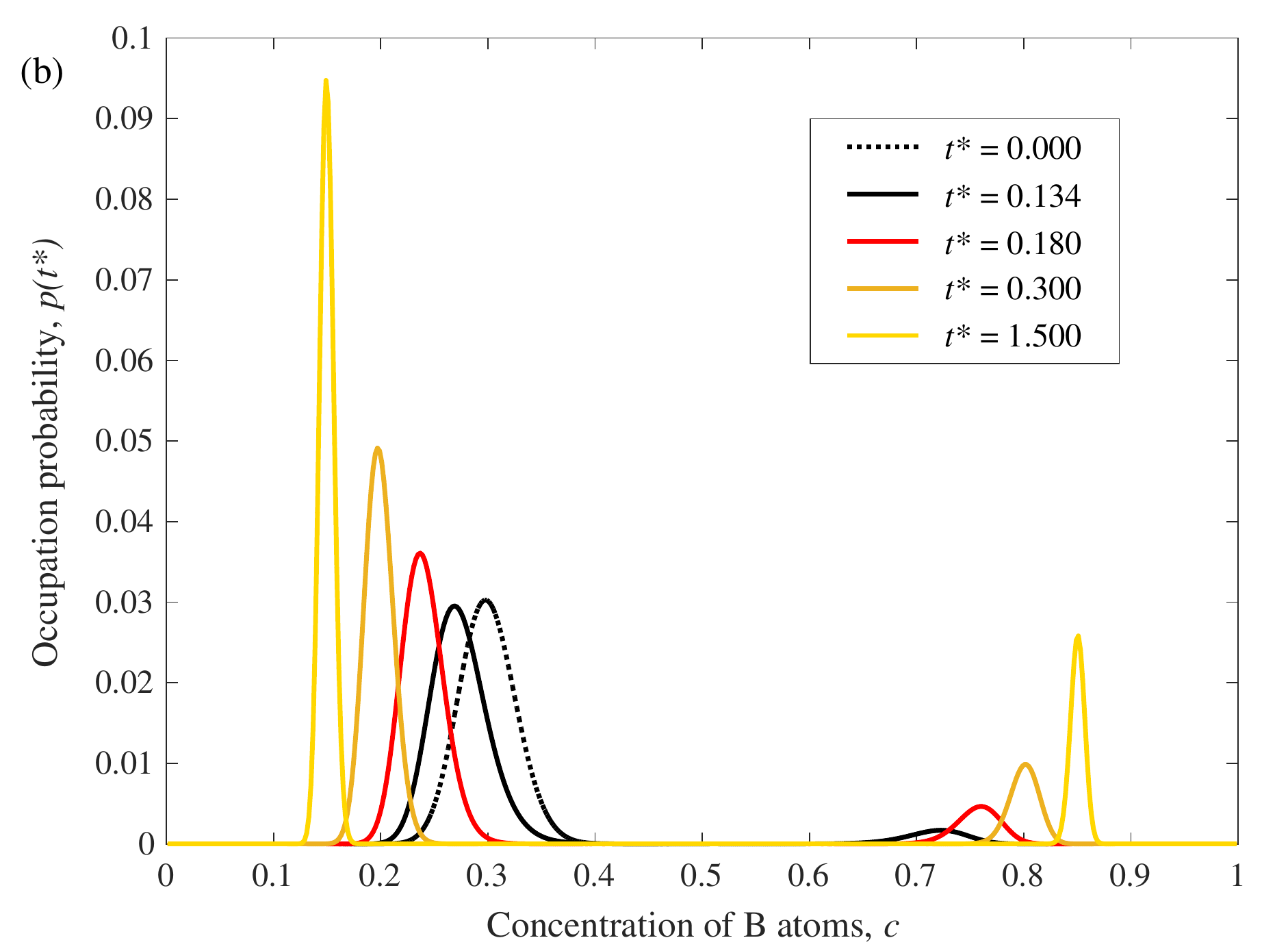}
\caption{\label{fig5:kinetic_separation_N0_1000} The calculated phase separation processes in (a) A--40.0\;at.\%\;B and (b) A--30.0\;at.\%\;B alloy systems at $T^{*}_R=0.20$ using $T^*_0=0.30$, $N=10^4$, and $N_0=10^3$. The time, $t$, is normalized by the relaxation time, $\tau$, as $t^*=t/\tau$. }
\end{figure}

Figure\;\ref{fig5:kinetic_separation_N0_1000}\;(a) shows the occupancy probabilities as a function of $c$ at five different times (expressed as a dimensionless ratio of time to the relaxation time, $t^*=t/\tau$) in a A--40.0\;at.\%\;B alloy. From the phase diagram in Fig.\;\ref{fig5:phase_diagram_phase_separation}, quenching this alloy from $T^*_0=0.30$ to $T^*_R=0.20$ falls within the spinodal limits and should thus lead to a continuous transformation.  The dotted curve in Fig.\;\ref{fig5:kinetic_separation_N0_1000}\;(a) represents the initial occupancy probability distribution at the high temperature, $T^*_0=0.30$.  As time increases, the occupancy probability evolves from the dashed distribution into two peaks (one at a dilute concentration of B and the other at a rich concentration) that eventually at $t^*=3.0$ correspond to the compositions of the two equilibrium solid-solutions at the temperature of the reservoir, $T^*_R=0.20$. At early times, the probability distribution between the two peaks of the evolving phases is non-zero --- this is a signature of a continuous transformation. There is a finite probability of finding any concentration between those of the two developing phases.  

A contrasting example is shown in Fig.\;\ref{fig5:kinetic_separation_N0_1000}\;(b), which shows the equivalent heat treatment for a A--30.0\;at.\%\;B alloy.  In this comparatively dilute alloy, the same thermal cycle places the alloy very close to the spinodal limit at the annealing (or reservoir) temperature. In this case, the initial probability distribution in Fig.\;\ref{fig5:kinetic_separation_N0_1000}\;(b) shifts to more dilute concentrations with time, and a new phase suddenly appears at high concentrations. The occupation probabilities of atomic configurations between the dilute and high concentrations are zero  --- this behavior is a signature of a discontinuous transformation (a nucleation and growth mechanism).  The B-rich phase with concentrations in the range $0.65 < c < 0.8$ appears from the initial distribution, but there are no occupied probabilities between $c=0.4$ and $c=0.65$.

Considering the influence of alloy composition, as the B concentration in the alloy increases from $c=0.3$ (Fig.\;\ref{fig5:kinetic_separation_N0_1000}\;(b)) to $c=0.4$ (Fig.\;\ref{fig5:kinetic_separation_N0_1000}\;(a)), the transformation mechanism switches from discontinuous to continuous. This transition is consistent with conventional wisdom in that the driving force for transformation increases with $c$ at the annealing temperature and has the effect of lowering the barrier to nucleation. Although not shown, it also was confirmed that the kinetic path is sensitive in a similar fashion to the annealing temperature: lowering the annealing temperature increases the driving force for decomposition and as a result shifts the mechanism from a discontinuous transformation path at high annealing temperatures to a continuous transformation path at lower annealing temperatures. 

It is worth noting that the equation of motion is a system of $R$ first-order, ordinary differential equations ($R$ is the number of energy eigenlevels). From a computational standpoint, these are relatively easy to solve.  For the system considered here ($R=500$ and $N=10^4$), the kinetic path from the initial state to stable equilibrium can be calculated in a few minutes on a laptop computer with 8\;GB of memory. This is an added advantage of the SEAQT approach when compared to other methods (e.g., kinetic Monte Carlo), where extensive information on particles and possible paths is required at each time-step. 

\subsection{\label{chap5_sec:level3_2}Estimated spinodal curves}
\begin{figure}
\begin{center}
\includegraphics[scale=0.42]{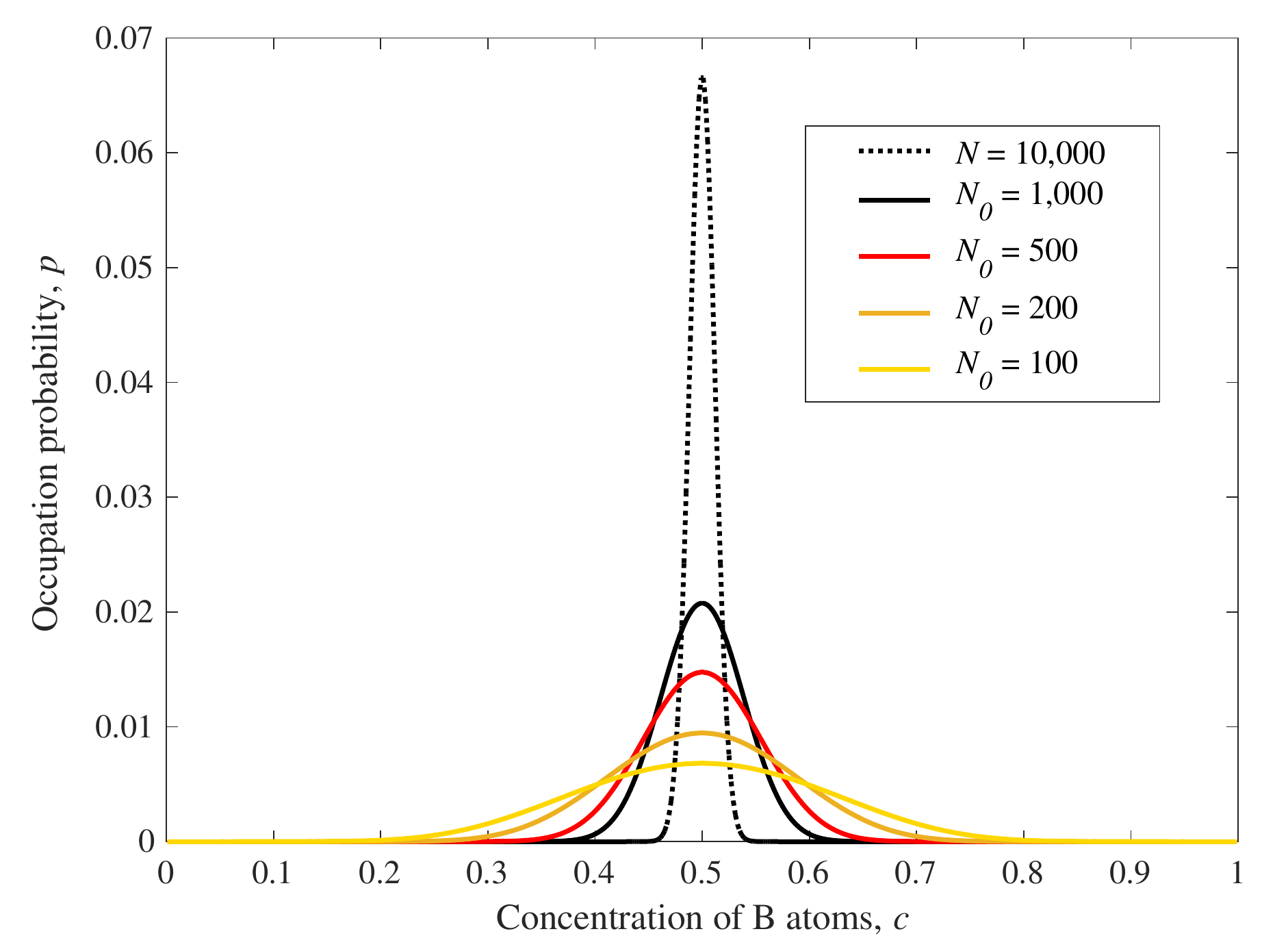}
\caption{\label{fig5:effects_of_Number_of_particles} The calculated initial probability distributions in a A--50.0\;at.\%\;B alloy system at $T^*_0=0.30$ using Eq.\;(\ref{eq5:grand_canonical_distribution}) with $N_0=1000$, $500$, $200$, and $100$. Here, an occupation probability calculated using $N_0=N=10^4$ is shown together as a dotted line. }
\end{center}
\end{figure}

Of course, being an initial value problem, the kinetic path is sensitive to the initial probability distribution. When an initial probability distribution, $p_j^0$, is prepared using a smaller $N_0$ (which corresponds to an initial state further from stable equilibrium at the initial temperature, $T^*_0$), the transformation path changes.  The effect of system size can be seen from Fig.\;\ref{fig5:effects_of_Number_of_particles}, where the initial probability distributions for systems with sizes, $N_0=1000$, $500$, $200$, and $100$, are calculated with Eq.\;(\ref{eq5:grand_canonical_distribution}) for an A--50.0\;at.\%\;B alloy at $T^*_0=0.30$. The larger the $N_0$ used to prepare the initial state, the sharper the peak in the occupancy probability distribution. In the limit of large $N_0$, the distribution is a delta function (at the most probable state of statistical mechanics). 

\begin{figure}
\begin{center}
\includegraphics[scale=0.12]{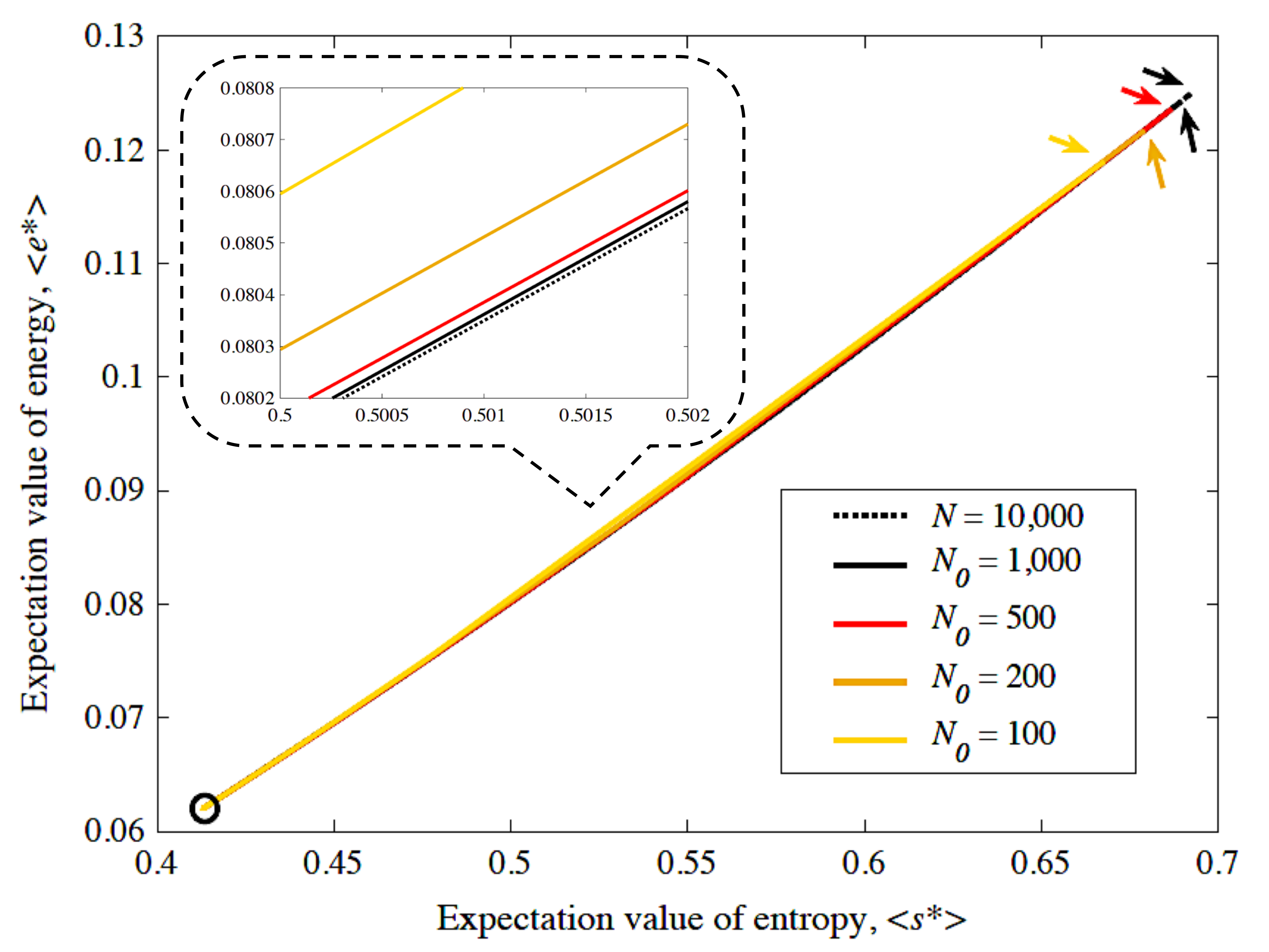}
\caption{\label{fig5:E_S_diagram_kinetic_pathway} The kinetic pathways of the phase separation process calculated with the SEAQT model using the initial probability distributions shown in Fig.\;\ref{fig5:effects_of_Number_of_particles} with $N=10^4$ (A--50.0\;at.\%\;B alloy with $T^{*}_0=0.30$ and $T^{*}_R=0.20$). 
The initial states of each path are indicated by arrows and the final states are shown by an open circle. The specific energy and entropy are normalized and denoted as $e^*$ and $s^*$, respectively.  }
\end{center}
\end{figure}

The kinetic pathways the system follows from the initial probability distributions of Fig.\;\ref{fig5:effects_of_Number_of_particles} are shown in Fig.\;\ref{fig5:E_S_diagram_kinetic_pathway}, where the kinetic path calculated with $N_0=N=10^4$ 
is shown as a dotted line. As seen from the enlarged inset in the figure, the deviation from the curve for $N_0=N=10^4$ becomes more significant as the initial fluctuation becomes larger. Note that although the initial states of each kinetic path in the energy-entropy diagram (Fig.\;\ref{fig5:E_S_diagram_kinetic_pathway}) are different, the final states of the paths correspond to the same stable equilibrium state since in each case the final state is one in which the alloy system is in mutual stable equilibrium with the same reservoir. 

\begin{figure}
\begin{center}
\includegraphics[scale=0.24]{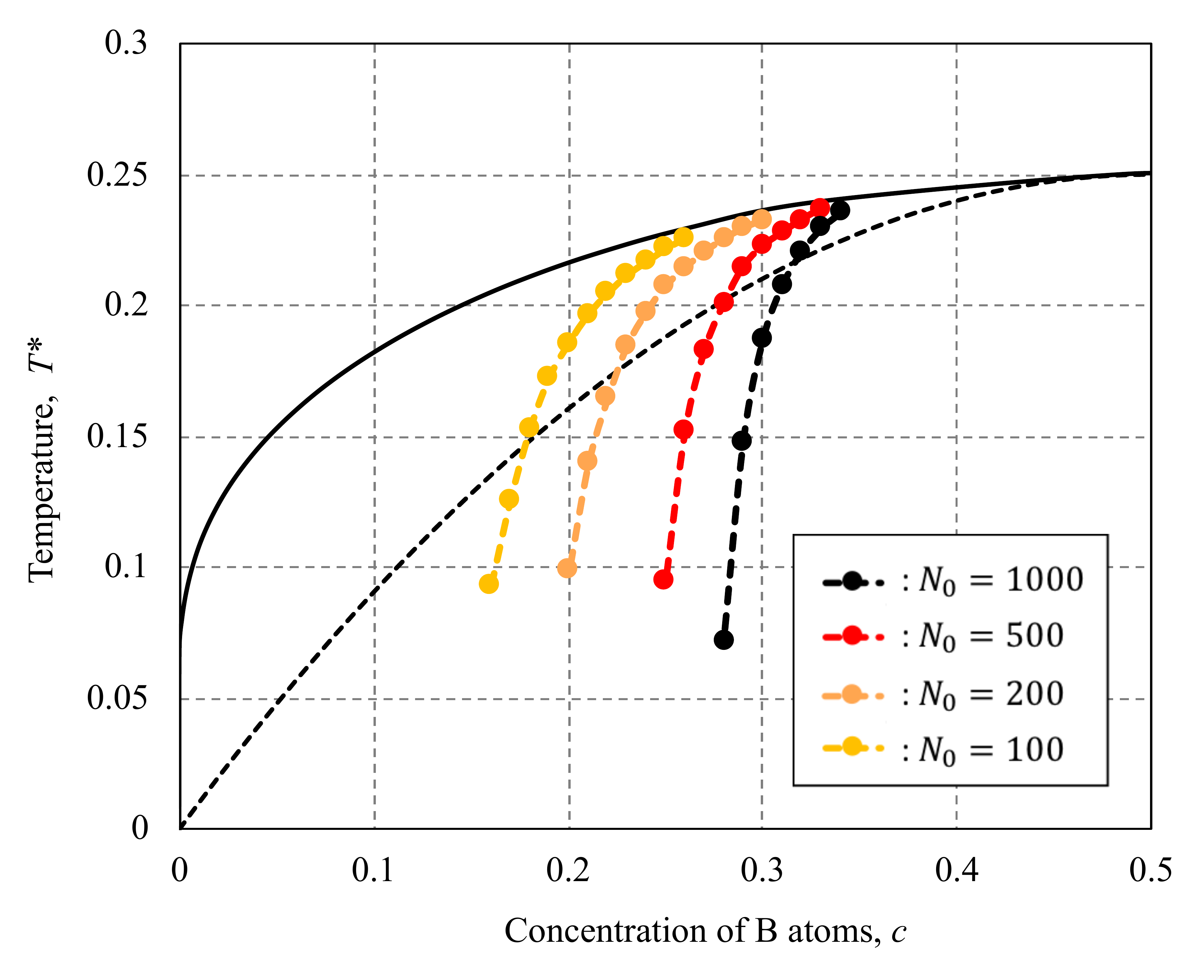}
\caption{\label{fig5:estimated_spinodal_line} The estimated spinodal curves using $T^*_0=0.30$ with the different initial probability distributions, $N_0=1000$, $500$, $200$, and $100$. When $T^*_R$ is inside/outside the spinodal curve, the transformation shows a continuous/discontinuous behavior. The solvus line (solid black line) and the spinodal curve (broken black line), which is determined from the free-energy analysis, are also shown together (Fig.\;\ref{fig5:phase_diagram_phase_separation}). }
\end{center}
\end{figure}

The fact that the initial state can affect the kinetic path has an interesting implication when it comes to representing the spinodal limit. When a phase decomposition process is continuous (spinodal in the present example), there is a non-zero occupation probability between the concentrations associated with the two stable concentration peaks during decomposition. On the other hand, when the transformation is discontinuous, there is a concentration range over which the occupation probabilities are zero when the second phase (precipitate) appears. Therefore, a spinodal curve can be determined by checking if occupation probabilities are zero or not in the concentration range between two peaks during decomposition process. In a numerical calculation, however, the probabilities have finite non-zero values even when those values are close to zero (e.g., $10^{-20}$). Practically speaking, we can select an arbitrary value, say, $10^{-5}$, as a cutoff below which the occupation probability is taken to be effectively zero to distinguish discontinuous occupation probabilities from continuous (non-zero) values. That is, when the second phase emerges and all probabilities between two peaks in the occupation probabilities are below $10^{-5}$, the transformation is taken to be discontinuous. Spinodal curves calculated from this criterion are shown in Fig.\;\ref{fig5:estimated_spinodal_line}. These spinodal curves are clearly sensitive to the initial state of the alloy system and are quite different from those determined from a free-energy analysis (the second derivative of the free-energy versus $c$ curve). This indicates that the onset of a continuous transformation is not simply a matter of the thermodynamic driving force at the transformation temperature. Instead, it also depends upon the initial state. Note that the criteria used for the distinction between continuous and discontinuous transformations, i.e., $10^{-5}$ here, should depend on the number of intervals in the concentration of B atoms, $R$ (see Sec.\;\ref{chap5_sec:level2_2}). When a larger value of $R$ is used, the criteria should be changed to a smaller value (here $R=500$ is used for the calculations). 
 
\subsection{\label{chap5_sec:level3_3}Scaling to dimensional time}
In the results shown in Fig.\;\ref{fig5:kinetic_separation_N0_1000}, the times, $t^*$, represent a dimensionless ratio of the actual dimensional time, $t$, and the relaxation time, $\tau$, from the SEAQT equation of motion. The latter represents a variable that tracks the dynamic progress from the initial state to the final equilibrium state.  The dimensional time can be extracted from $t^*$ through a comparison with experimental data \cite{beretta2017steepest,li2018multiscale} or from $ab$ $initio$ calculations \cite{beretta2014steepest,li2016generalized,li2018steepest,yamada2018method}. 

While SEAQT framework predicts the transformation mechanism (nucleation-growth or spinodal decomposition) for a given eigenstructure by selecting the path from the initial state with the steepest entropy ascent principle, the actual time required to traverse this path depends upon the rate of entropy production associated with the unit processes. For a nucleation process involving the assembly of subcritical embryos, entropy production is much slower than for the diffusion throughout a spinodally decomposing material. Thus, the scaling that maps the relaxation time, $\tau$, to dimensional time should be different for the nucleation-growth and spinodal mechanisms.  

Here, the dimensional time dependence is extracted via comparisons of the relaxation time to experimental transformation kinetics from the Co--Cu alloy system. 
The Cu--Co system has a positive mixing enthalpy (positive $V(\bm{0})$ in Eq.\;(\ref{eq5:total_energy_phase_separation})) and a large miscibility gap extending over almost the whole concentration range (see the phase diagram in reference \cite{nishizawa1984co}). The discontinuous transformation mechanism (nucleation-growth) has been investigated extensively in the Cu-rich region (Cu--0.5$\sim$2.7\;at.\%\;Co alloys) \cite{legoues1984influence,wendt1985atom,hattenhauer1993decomposition}, and the continuous transformation mechanism (spinodal decomposition) has been observed in Cu--10\;at.\%\;Co alloy at 713\;K \cite{busch1996high}. 

\begin{figure}
\begin{center}
\includegraphics[scale=0.30]{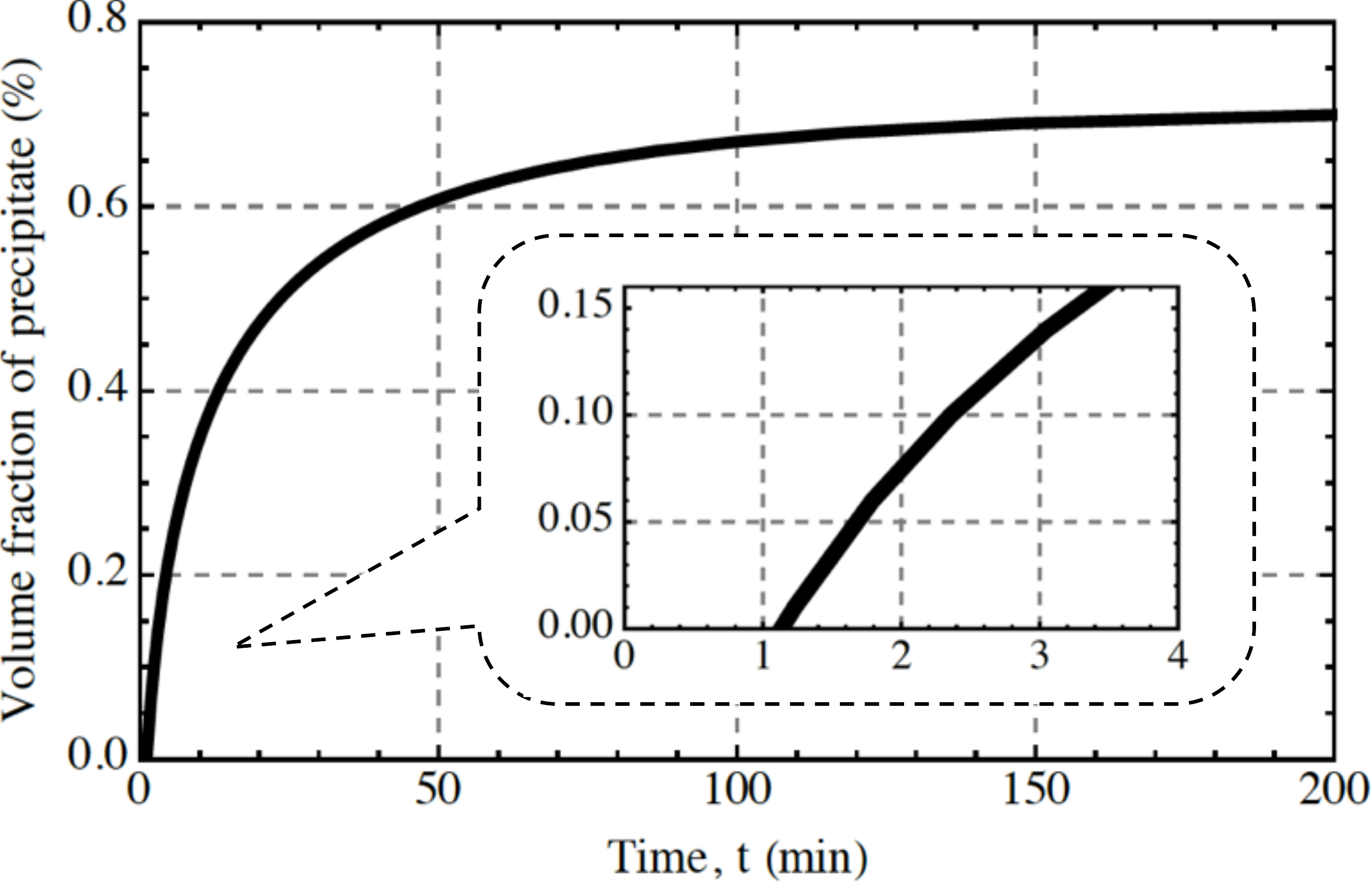}
\caption
{\label{fig5:kinetic_separation_realtime_C_1B} The dimensional time dependence of the precipitate volume fraction during nucleation and growth in Cu--1.0\;at.\%\;Co annealed at 823\;K calculated with SEAQT using $T^{*}_R=0.089$, $T^{*}_0=0.30$, $N=10^4$, and $N_0=10^2$. The relaxation time is correlated with the experimental kinetics of Cu--1\;at.\%\;Co alloy annealed at 823\;K \cite{hattenhauer1993decomposition}. The inset has a time range of 0-4\;min and the incubation period for the nucleation process obtained from the intercept with the abscissa is approximately 1.2\;min.  }
\end{center}
\end{figure}

\begin{figure}
\begin{center}
\includegraphics[scale=0.54]{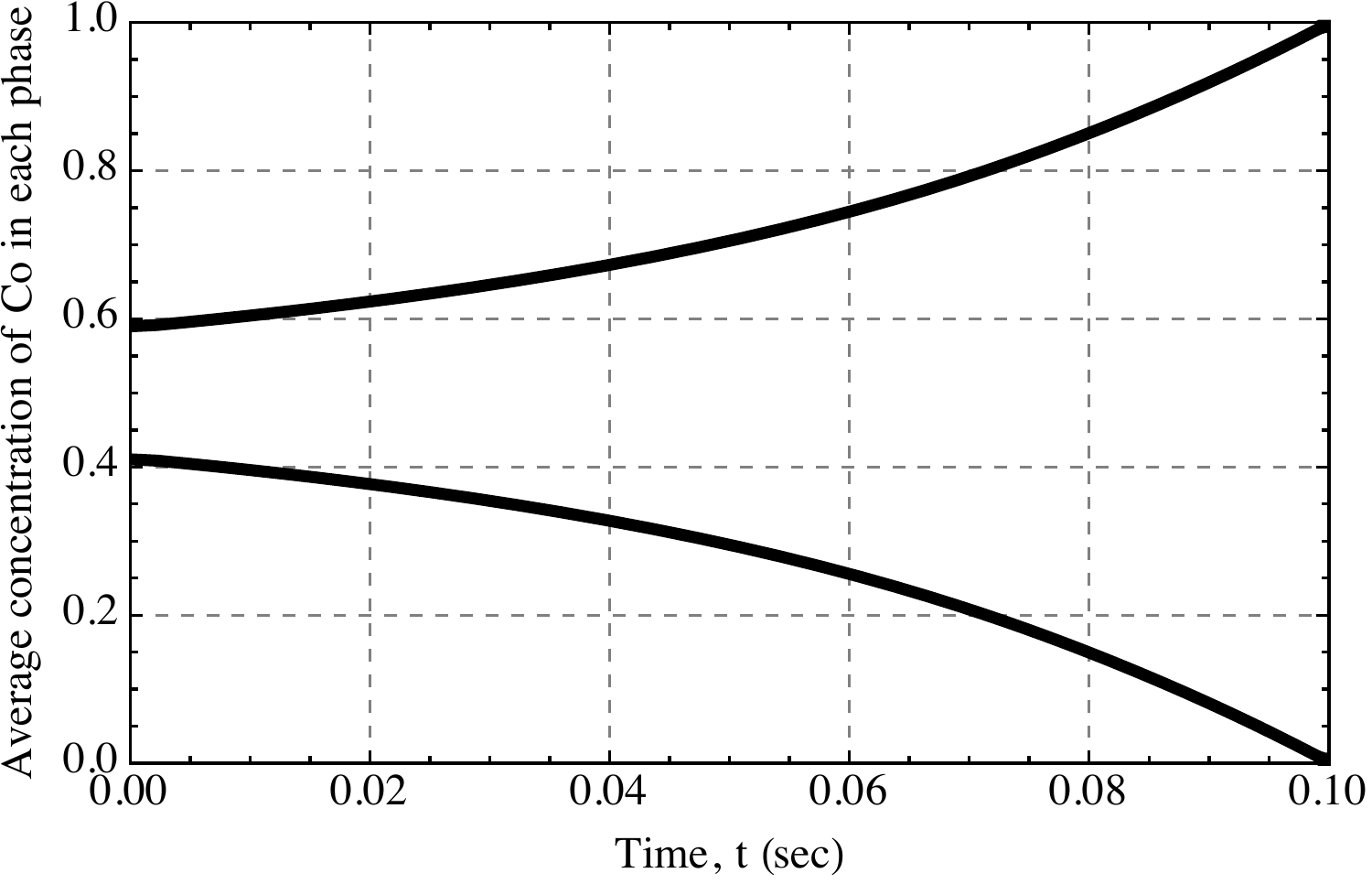}
\caption
{\label{fig5:kinetic_separation_realtime_C_50B} The dimensional time dependence of the Co concentration in the Cu- and Co-rich phases during spinodal decomposition in Cu--50.0\;at.\%\;Co annealed at 823\;K calculated with SEAQT using $T^{*}_R=0.089$, $T^{*}_0=0.30$, $N=10^4$, and $N_0=10^2$. The relaxation time is correlated with the experimental diffusion coefficient \cite{dohl1984measurement} and the characteristic wave length of the spinodal microstructure \cite{liu1993spinodal}.}
\end{center}
\end{figure}

The procedures for scaling the dimensional time to the relaxation time for each transformation mechanism (nucleation-growth and spinodal decomposition) are shown in Appendices\;\ref{chap5_sec:level5_1} and \ref{chap5_sec:level5_2}, respectively.  After scaling the relaxation time, $\tau$, to experimental data, the calculated kinetics from the SEAQT framework can be presented in terms of dimensional time. Figures\;\ref{fig5:kinetic_separation_realtime_C_1B} and \ref{fig5:kinetic_separation_realtime_C_50B} show the time-dependence of the nucleated precipitate volume fraction (the Co-rich phase) and the concentration of Co atoms in the spinodal decomposed phases, respectively. The predicted time-evolution processes show opposite tendencies: the speed of the transformation slows as nucleation and growth proceeds, whereas spinodal decomposition is predicted to accelerate as the transformation proceeds. Thus, the different experimental scalings for $\tau$ make it possible to place nucleation-growth and spinodal decomposition on very different dimensional time scales: spinodal decomposition is scaled to times less than a second whereas nucleation-growth extends over a period of 2 or 3\;hours.  

\section{\label{chap5_sec:level4}Conclusions}
The quantum mechanical-based SEAQT framework was applied to the decomposition of a binary solid-solution using a pseudo-eigenstructure based on the mean-field approximation. In this Part\;I, the different behaviors of continuous and discontinuous transformations are explored. It is confirmed that the SEAQT approach is able to predict the transformation characteristic of continuous and discontinuous transformation mechanisms. The kinetic path is sensitive to the initial state of the alloy and the annealing temperature, and the spinodal limits estimated from the SEAQT model show some quantitative difference from the conventional spinodal limit calculated from a free-energy analysis. Furthermore, very different dimensional time dependencies of the continuous and discontinuous transformation mechanisms are readily obtained by calibrating the SEAQT relaxation time to experimental spinodal data and nucleation-growth data. 

It is noteworthy that the SEAQT model with a mean-field approximation is computationally efficient. Kinetic paths from an initial state to stable equilibrium in a system considered here were obtained in minutes on a standard laptop computer.

\section*{ACKNOWLEDGEMENT}
We acknowledge the National Science Foundation (NSF) for support through Grant DMR-1506936. \\

\begin{appendices}

\section*{\label{chap5_sec:level5}Appendix}

\section{\label{chap5_sec:level5_1}Scaling to dimensional time for nucleation-growth}
The nucleation-growth mechanism has been investigated in the Cu--Co alloy system \cite{legoues1984influence,wendt1985atom,hattenhauer1993decomposition}. The relaxation time can be related to the dimensional time, $t$, in the calculated discontinuous phase transformation using experimental data for a Cu--1\;at.\%\;Co alloy isothermally aged at 823\;K \cite{hattenhauer1993decomposition}. 

\begin{figure}
\begin{center}
\includegraphics[scale=0.60]{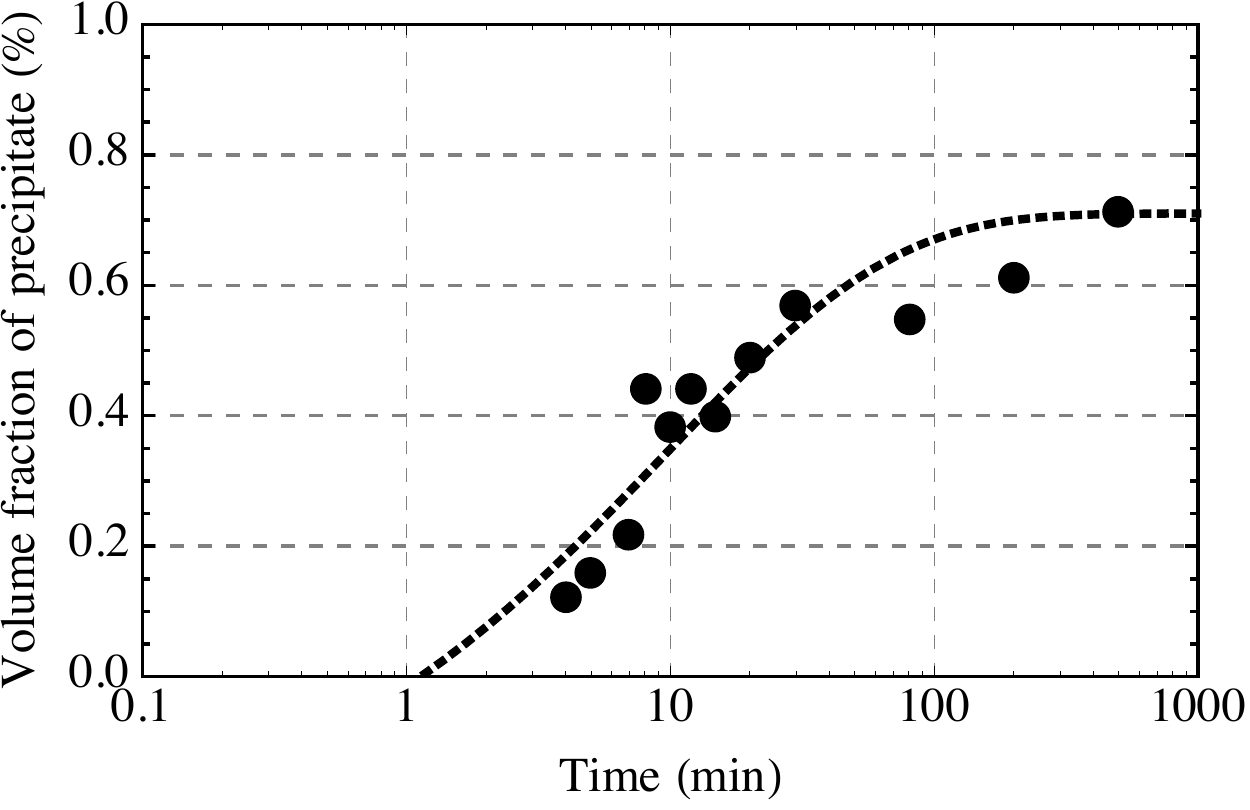}
\caption
{\label{fig5:precipitated_volume_fraction_experiments} The experimentally measured volume fraction of precipitate (or Co-rich phase) in a Cu--1\;at.\%\;Co alloy isothermally aged at 823\;K. The black circles are the original data \cite{hattenhauer1993decomposition} and the dotted line is the fitting function, $f_p = f_p^{\mbox{\scriptsize max}} - e^{-K t^n}$, where $f_p^{\mbox{\scriptsize max}}=0.71$, $K=0.3217$, and $n=0.5004$.} 
\end{center}
\end{figure}

The measured data of the precipitated volume fraction at Cu--1\;at.\%\;Co alloy at 823\;K is shown in Fig.\;\ref{fig5:precipitated_volume_fraction_experiments}, where the following fitting function is shown as well:
\begin{equation}
f_p = f_p^{\mbox{\scriptsize max}} - e^{-K t^n}  \;,     \label{eq5:fitting_function_nucleation_growth}
\end{equation}
where $f_p$ is the volume fraction of the precipitate, $f_p^{\mbox{\scriptsize max}}$ is the maximum measured value of $f_p$, $t$ is the annealing time, and $K$ and $n$ are the fitting parameters. Equation\;(\ref{eq5:fitting_function_nucleation_growth}) is rewritten as  
\begin{equation}
t = \left[ - \frac{1}{K} \mathrm{ln} (f_p^{\mbox{\scriptsize max}} - f_p) \right] ^{\frac{1}{n}}  \; .    \label{eq5:fitting_function_nucleation_growth_t}
\end{equation}
The annealing time, $t$, can be determined once the volume fraction, $f_p$, is known at each time. 

\begin{figure}
\begin{center}
\includegraphics[scale=0.55]{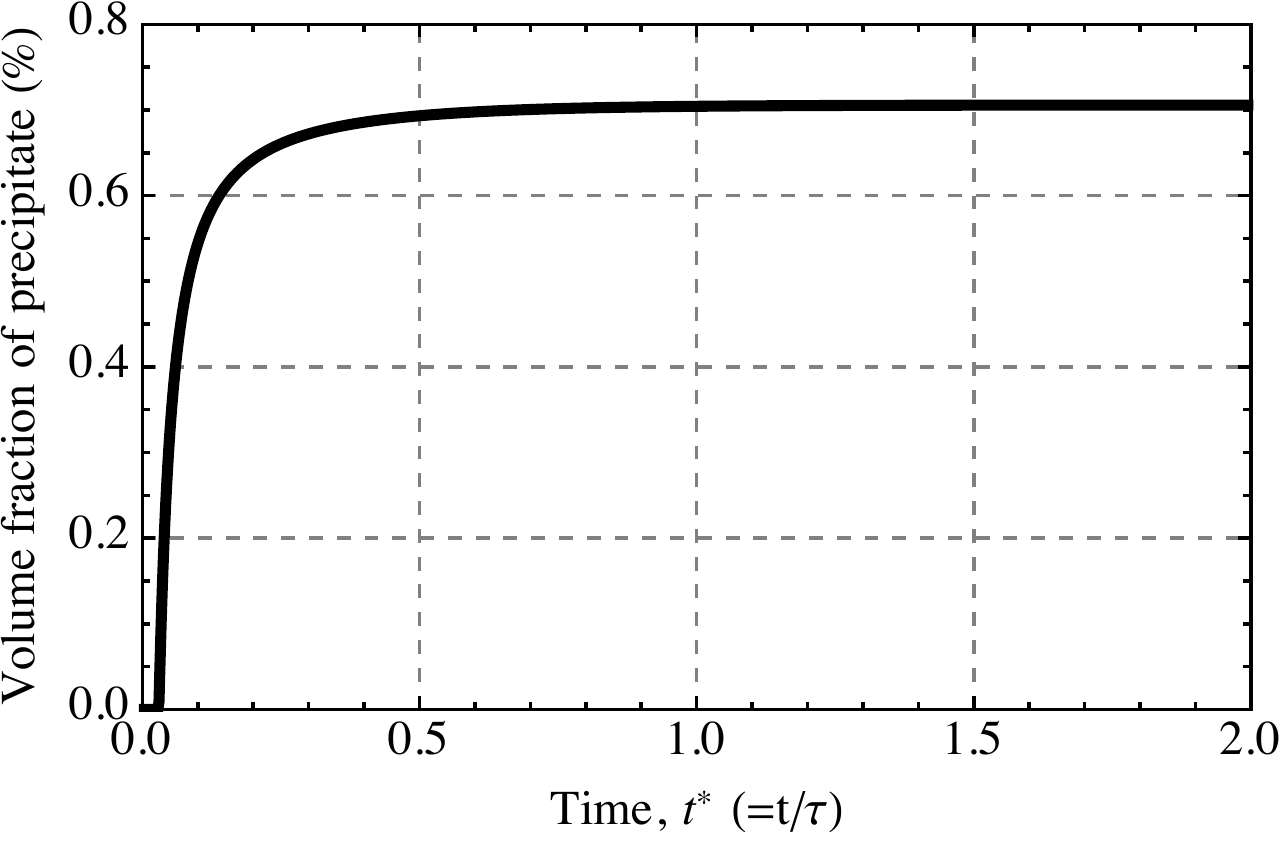}
\caption{\label{fig5:volume_fraction_precipitate_SEAQT} The time dependences of the volume fraction of precipitate (or B-rich phase) in a A--1.0\;at.\%\;B alloy system  calculated with the SEAQT modeling using $T^{*}_R=0.089$, $T^{*}_0=0.30$, $N=10^4$, and $N_0=10^2$.  }  
\end{center}
\end{figure}

\begin{figure}
\begin{center}
\includegraphics[scale=0.55]{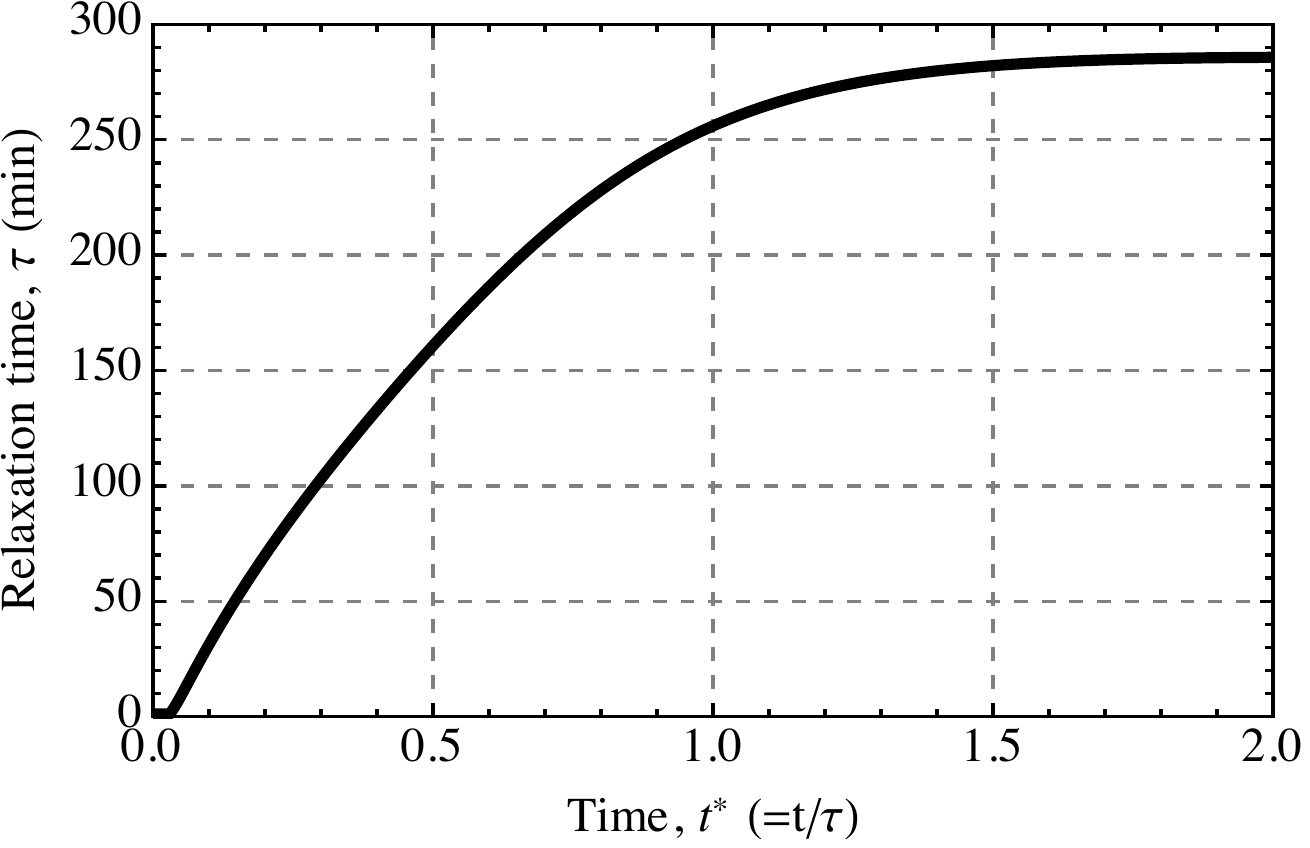}
\caption{\label{fig5:relaxation_time_nucleation_mechanism} The time dependence of the relaxation time, $\tau$, in a Cu--1.0\;at.\%\;Co alloy system when a sample with some initial concentration fluctuations is annealed at $823$\;K. It is estimated using Eq.\;(\ref{eq5:fitting_function_nucleation_growth_t}) with the result shown in Fig.\;\ref{fig5:volume_fraction_precipitate_SEAQT}.   }  
\end{center}
\end{figure}

Although the real temperatures of the calculated phase diagram (shown in Fig.\;\ref{fig5:phase_diagram_phase_separation}) were estimated using the reported regular solution parameter, $\Omega=V(\bm{0})/2=33,300$ (J/mol) \cite{hattenhauer1993decomposition}, the phase diagram had some differences with the experimentally determined one \cite{nishizawa1984co}. For this reason, the normalized temperature, which corresponds to 823\;K, is found by searching for the condition for which the calculated $f_p^{\mbox{\scriptsize max}}$ becomes 0.71. Since $f_p^{\mbox{\scriptsize max}} \sim 0.71$ at $T^{*}_R=0.089$, the normalized annealing temperature, $T^{*}_R=0.089$, is used here for the calculation. The calculated time dependence of the volume fraction of precipitate, $f_p$, predicted by SEAQT is shown in Fig.\;\ref{fig5:volume_fraction_precipitate_SEAQT}. The determined time dependence of the relaxation time, $\tau$, is shown in Fig.\;\ref{fig5:relaxation_time_nucleation_mechanism}. Note that Eq.\;(\ref{eq5:fitting_function_nucleation_growth}) has negative values below $t \approx 1$ (see Fig.\;\ref{fig5:precipitated_volume_fraction_experiments}), but this does not cause difficulties when determining the relaxation time.

\section{\label{chap5_sec:level5_2}Scaling to dimensional time for spinodal decomposition}
\begin{figure}
\begin{center}
\includegraphics[scale=0.45]{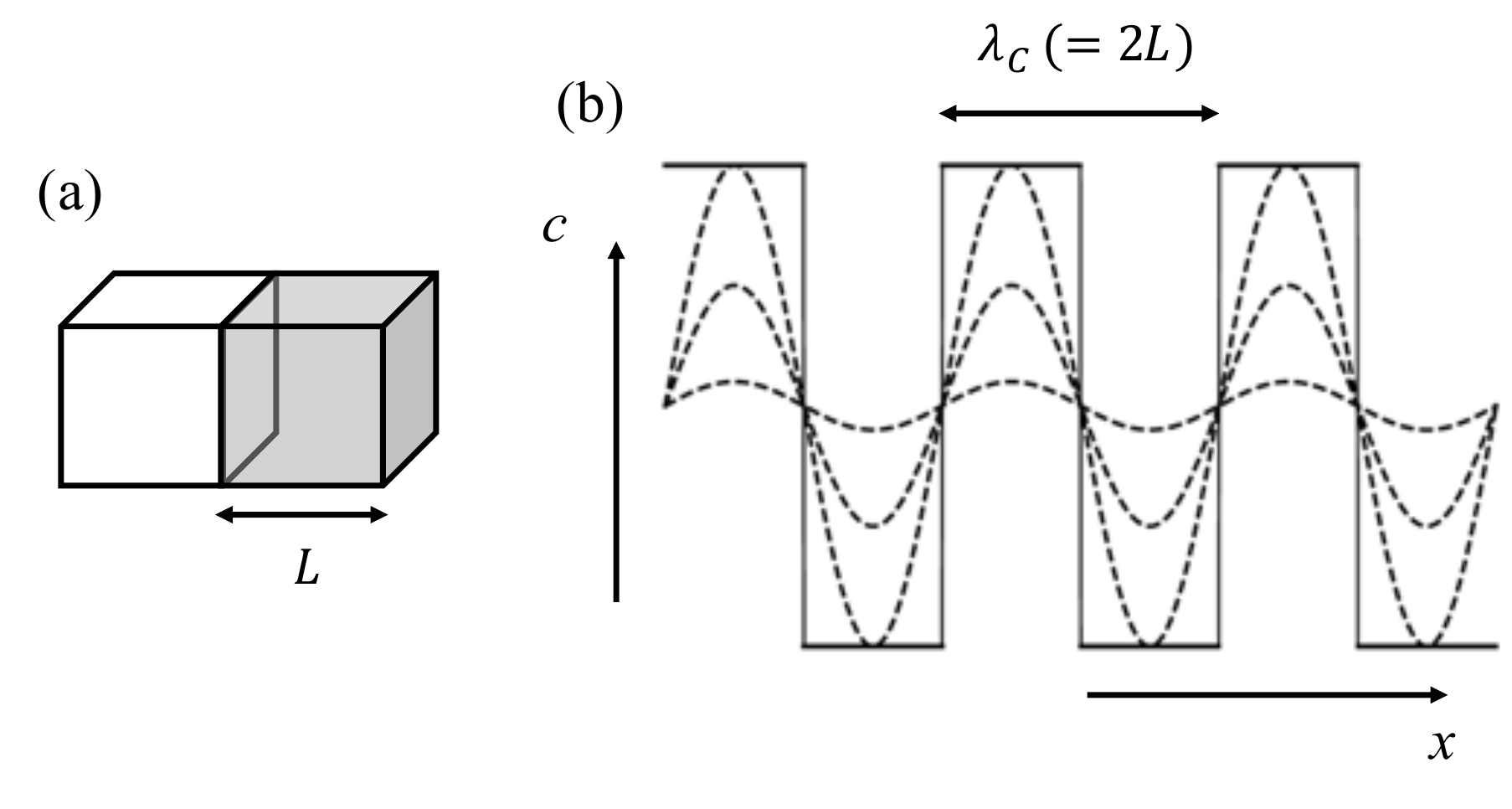}
\caption{\label{fig5:atomic_diffusion_pic_cubic_spinodal} (a) One dimensional atomic diffusion between assumed cube-shaped phases with side length, $L$ (each phase corresponds to either $\alpha$- or $\beta$-phase). (b) the schematic time-evolution process of spinodal decomposition; the broken lines are part way through the evolution process, and the solid lines are the final distribution. The side length of the cube-shaped regions shown in (a) would correspond to half of the characteristic wave length of the spinodal microstructure, $\lambda _c$; i.e., $L = \lambda _c /2$.   } 
\end{center}
\end{figure}

To scale the relaxation time, $\tau$, to dimensional time for a continuous transformation, the reported diffusion coefficient \cite{dohl1984measurement} and the characteristic wave length of the spinodal microstructure \cite{liu1993spinodal} in a Cu--Co alloy system are used.  Atomic diffusion is assumed between the cube-shaped A-rich ($\alpha$) and B-rich ($\beta$) phases in a A--50.0\;at.\%\;B alloy system, where the edge length of the phases, $L$, corresponds to half the characteristic wave length of the spinodal microstructure, $\lambda_c$ (see Fig.\;\ref{fig5:atomic_diffusion_pic_cubic_spinodal}). The diffusion equation for a constant diffusivity is given by 
\begin{equation}
\frac{\partial c^{\alpha / \beta}}{\partial t} = D \nabla^2 c^ {\alpha / \beta}   \;,    \label{eq5:diffusion_equation}
\end{equation}
where $D$ is the diffusion coefficient and $c^ {\alpha / \beta}$ is the concentration of B-type atoms in the $\alpha/\beta$-phase. The Laplacian can be replaced by expressing the concentration on each of the six surfaces of the cube as a Taylor series expanded about $c^ {\alpha / \beta}$ at the cube center, $c^ {\alpha / \beta}_0$, and summing the series (up to the quadratic terms). With this approximation, Eq.\;(\ref{eq5:diffusion_equation}) becomes
\begin{equation}
\frac{\partial c^{\alpha / \beta}}{\partial t} \approx D \frac{6}{(L/2)^2} (c^{\beta / \alpha} - c^{\alpha / \beta}_0)     \;,    \label{eq5:diffusion_equation_approximation}
\end{equation}
where $L$ is the edge length of the cube-shaped phases and given as $L=\lambda_c/2$. When an average quantity of concentration of B-type atoms in each phase, $\langle c \rangle^{\alpha / \beta}$, is taken, Eq.\;(\ref{eq5:diffusion_equation_approximation}) is written as
\begin{equation}
\frac{\partial \langle c \rangle^{\alpha / \beta}}{\partial t} = D \frac{6}{(L/2)^2} (\langle c \rangle^{\beta / \alpha} - \langle c \rangle^{\alpha / \beta})     \;.    \label{eq5:diffusion_equation_approximation_average}
\end{equation}

For the equivalent SEAQT system, the concentration change rate is given as 
\begin{equation}
\frac{\partial \langle c \rangle^{\alpha / \beta}}{\partial t} \Rightarrow \frac{d \langle c \rangle^{\alpha / \beta}}{d t^*}    \;,     \label{eq5:dc_dt_SEAQT}
\end{equation}
where $t^*$ is a normalized time ($t^*=t/\tau$). Thus, the relaxation time, $\tau$, is derived as
\begin{equation}
\tau=\frac{(\lambda_c/2)^2}{24 D (\langle c \rangle^{\beta / \alpha} - \langle c \rangle^{\alpha / \beta})} \frac{d \langle c \rangle^{\alpha / \beta}}{d t^*}  \;.    \label{eq5:relaxation_time_spinodal}
\end{equation}
Note that $\langle c \rangle^{\alpha / \beta}$ is a function of time, and $D$ is also time-dependent because the temperature in an alloy system changes with time. Here, however, it is assumed that $D$ is time-independent and the value used for $D$ is that at the annealing temperature. 

\begin{figure}
\begin{center}
\includegraphics[scale=0.6]{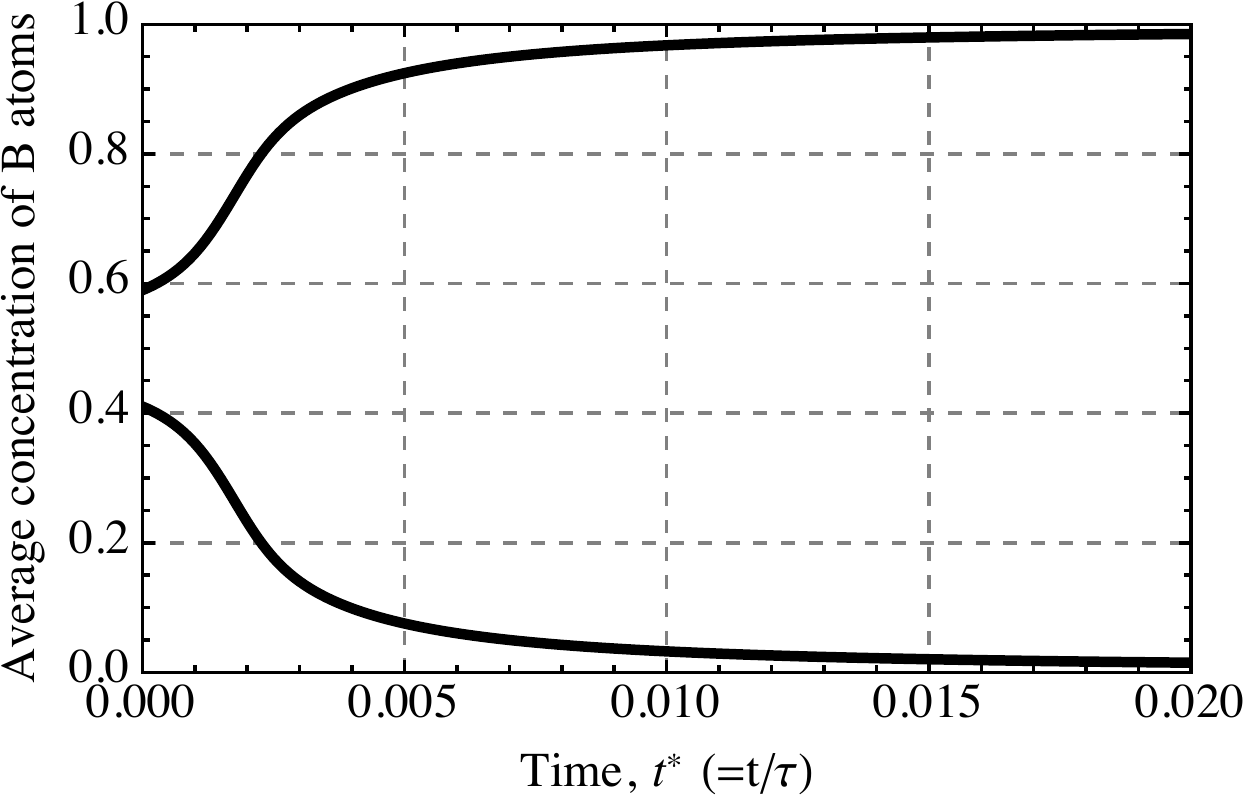}
\caption{\label{fig5:average_concB_each_phase} The time dependence of average concentration of B-type atoms in A-rich ($\alpha$) and B-rich ($\beta$) phases calculated with the SEAQT model using $T^{*}_R=0.089$, $T^{*}_0=0.30$, $N=10^4$, and $N_0=10^2$. The averages are, respectively, taken from the calculated occupation probabilities in the concentration ranges 0$\sim$50\;at.\%\;B and 50$\sim$100\;at.\%\;B.  } 
\end{center}
\end{figure} 

\begin{figure}
\begin{center}
\includegraphics[scale=0.6]{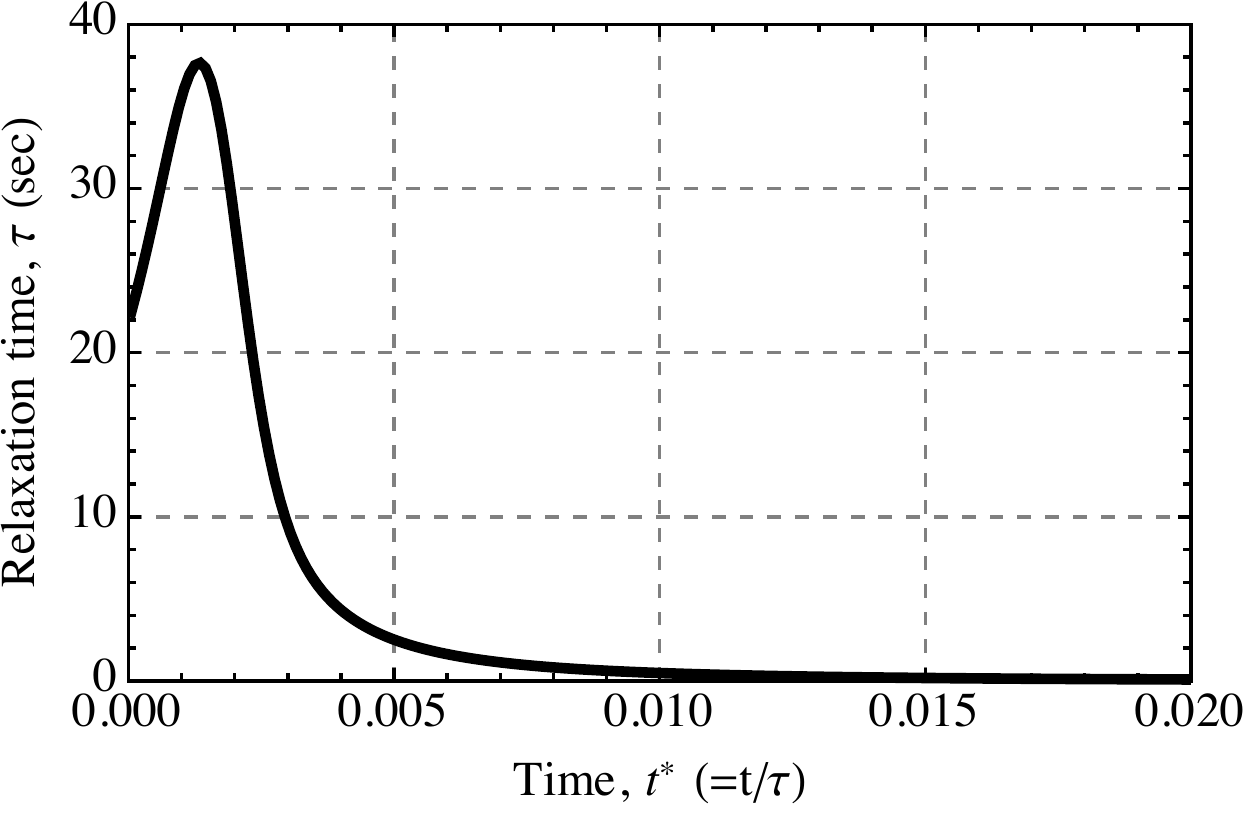}
\caption
{\label{fig5:relaxation_time_spinodal} The time dependence of the relaxation time, $\tau$, in a Cu--50.0\;at.\%\;Co alloy system when a sample with some initial concentration fluctuations is annealed at $823$\;K. It is estimated using Eq.\;(\ref{eq5:relaxation_time_spinodal}) with the result shown in Fig.\;\ref{fig5:average_concB_each_phase} and the reported experimental data \cite{dohl1984measurement,liu1993spinodal}, $D=0.43\; \mbox{exp}(- 2.22 \; \mbox{eV} /k_BT)$ and $\lambda_c \approx 3.5$\;nm.  }  
\end{center}
\end{figure} 

The experimental data of the diffusion coefficient and the characteristic spinodal wave length in a Cu--Co alloy system are, respectively, $D=0.43\;\mbox{exp}(- 2.22 \; \mbox{eV} /k_BT)$ (for Cu--0.1\;$\sim$\;0.15\;at.\%\;Co with 640\;$\sim$\;848\;K) \cite{dohl1984measurement} and $\lambda_c \approx 3.5$\;nm \cite{liu1993spinodal}. Since it is estimated that $T^{*}_R=0.089$ corresponds to 823\;K in Appendix\;\ref{chap5_sec:level5_1}, the spinodal decomposition behavior at 823\;K is investigated here for Cu--50.0\;at.\%\;Co alloy assuming that the diffusion coefficient is not sensitive to the composition. The calculated time dependence of the average concentration of B atoms in each phase using the SEAQT model is shown in Fig.\;\ref{fig5:average_concB_each_phase}, where the averages of each phase are, respectively, taken in the concentration ranges 0$\sim$50\;at.\%\;B and 50$\sim$100\;at.\%\;B. The determined time dependence of the relaxation time, $\tau$, using Eq.\;(\ref{eq5:relaxation_time_spinodal}) is shown in Fig.\;\ref{fig5:relaxation_time_spinodal}.

\end{appendices}

\bibliographystyle{ieeetr}
\bibliography{ref}

\begin{thebibliography}{10}

\bibitem{gibbs1906scientific}
J.~W. Gibbs, {\em The scientific papers of {J}. {W}illard {G}ibbs}, vol.~1.
\newblock Longmans, Green and Company, 1906.

\bibitem{cahn1961spinodal}
J.~W. Cahn, ``On spinodal decomposition,'' {\em Acta metallurgica}, vol.~9,
  no.~9, pp.~795--801, 1961.

\bibitem{balluffi2005kinetics}
R.~W. Balluffi, S.~M. Allen, and W.~C. Carter, {\em Kinetics of materials}.
\newblock John Wiley \& Sons, 2005.

\bibitem{schmelzer2004dynamics}
J.~W.~P. Schmelzer, A.~R. Gokhman, and V.~M. Fokin, ``Dynamics of first-order
  phase transitions in multicomponent systems: a new theoretical approach,''
  {\em Journal of Colloid and Interface Science}, vol.~272, no.~1,
  pp.~109--133, 2004.

\bibitem{soisson2006kinetic}
F.~Soisson, ``Kinetic {M}onte {C}arlo simulations of radiation induced
  segregation and precipitation,'' {\em Journal of Nuclear Materials},
  vol.~349, no.~3, pp.~235--250, 2006.

\bibitem{gao2018theoretical}
Y.~Gao, Y.~Zhang, D.~Schwen, C.~Jiang, C.~Sun, J.~Gan, and X.-M. Bai,
  ``Theoretical prediction and atomic kinetic {M}onte {C}arlo simulations of
  void superlattice self-organization under irradiation,'' {\em Scientific
  reports}, vol.~8, 2018.

\bibitem{hatsopoulos1976-I}
G.~N. Hatsopoulos and E.~P. Gyftopoulos, ``A unified quantum theory of
  mechanics and thermodynamics. {P}art {I}. postulates,'' {\em Foundations of
  Physics}, vol.~6, no.~1, pp.~15--31, 1976.

\bibitem{hatsopoulos1976-IIa}
G.~N. Hatsopoulos and E.~P. Gyftopoulos, ``A unified quantum theory of
  mechanics and thermodynamics. {P}art {II}a. {A}vailable energy,'' {\em
  Foundations of Physics}, vol.~6, no.~2, pp.~127--141, 1976.

\bibitem{hatsopoulos1976-IIb}
G.~N. Hatsopoulos and E.~P. Gyftopoulos, ``A unified quantum theory of
  mechanics and thermodynamics. {P}art {II}b. {S}table equilibrium states,''
  {\em Foundations of Physics}, vol.~6, no.~4, pp.~439--455, 1976.

\bibitem{hatsopoulos1976-III}
G.~N. Hatsopoulos and E.~P. Gyftopoulos, ``A unified quantum theory of
  mechanics and thermodynamics. {P}art {III}. {I}rreducible quantal
  dispersions,'' {\em Foundations of Physics}, vol.~6, no.~5, pp.~561--570,
  1976.

\bibitem{beretta2005generalPhD}
G.~P. Beretta, {\em On the general equation of motion of quantum thermodynamics
  and the distinction between quantal and nonquantal uncertainties}.
\newblock PhD thesis, Massachusetts Institute of Technology, 1981.

\bibitem{li2016steepest}
G.~Li and M.~R. von Spakovsky, ``{S}teepest-entropy-ascent quantum
  thermodynamic modeling of the relaxation process of isolated chemically
  reactive systems using density of states and the concept of hypoequilibrium
  state,'' {\em Physical Review E}, vol.~93, no.~1, p.~012137, 2016.

\bibitem{yamada2018method}
R.~Yamada, M.~R. von Spakovsky, and W.~T. Reynolds~Jr., ``A method for
  predicting non-equilibrium thermal expansion using steepest-entropy-ascent
  quantum thermodynamics,'' {\em Journal of Physics: Condensed Matter},
  vol.~30, no.~32, p.~325901, 2018.

\bibitem{beretta1985quantum}
G.~P. Beretta, E.~P. Gyftopoulos, and J.~L. Park, ``Quantum thermodynamics. {A}
  new equation of motion for a general quantum system,'' {\em Il Nuovo Cimento
  B}, vol.~87, no.~1, pp.~77--97, 1985.

\bibitem{beretta2006nonlinear}
G.~P. Beretta, ``{N}onlinear model dynamics for closed-system, constrained,
  maximal-entropy-generation relaxation by energy redistribution,'' {\em
  Physical Review E}, vol.~73, no.~2, p.~026113, 2006.

\bibitem{beretta2009nonlinear}
G.~P. Beretta, ``{N}onlinear quantum evolution equations to model irreversible
  adiabatic relaxation with maximal entropy production and other nonunitary
  processes,'' {\em Reports on Mathematical Physics}, vol.~64, no.~1/2,
  pp.~139--168, 2009.

\bibitem{gyftopoulos1997entropy}
E.~P. Gyftopoulos and E.~Cubukcu, ``Entropy: thermodynamic definition and
  quantum expression,'' {\em Physical Review E}, vol.~55, no.~4, p.~3851, 1997.

\bibitem{cubukcu1993thermodynamics}
E.~Cubukcu, {\em {T}hermodynamics as a non-statistical theory}.
\newblock PhD thesis, Massachusetts Institute of Technology, 1993.

\bibitem{yamada2018steepest}
R.~Yamada, M.~R. von Spakovsky, and W.~T. Reynolds~Jr.,
  ``{S}teepest-{E}ntropy-{A}scent {Q}uantum {T}hermodynamics {M}odels in
  {M}aterials {S}cience,'' {\em (preparing)}.

\bibitem{li2016generalized}
G.~Li and M.~R. von Spakovsky, ``Generalized thermodynamic relations for a
  system experiencing heat and mass diffusion in the far-from-equilibrium realm
  based on steepest entropy ascent,'' {\em Physical Review E}, vol.~94, no.~3,
  p.~032117, 2016.

\bibitem{li2016modeling}
G.~Li and M.~R. von Spakovsky, ``Modeling the nonequilibrium effects in a
  nonquasi-equilibrium thermodynamic cycle based on steepest entropy ascent and
  an isothermal-isobaric ensemble,'' {\em Energy}, vol.~115, pp.~498--512,
  2016.

\bibitem{li2017study}
G.~Li and M.~R. von Spakovsky, ``Study of {N}onequilibrium {S}ize and
  {C}oncentration {E}ffects on the {H}eat and {M}ass {D}iffusion of
  {I}ndistinguishable {P}articles using {S}teepest-{E}ntropy-{A}scent {Q}uantum
  {T}hermodynamics,'' {\em Journal of Heat Transfer}, vol.~139, no.~12,
  p.~122003, 2017.

\bibitem{li2016steepest2}
G.~Li and M.~R. von Spakovsky, ``Steepest-entropy-ascent quantum thermodynamic
  modeling of the far-from-equilibrium interactions between nonequilibrium
  systems of indistinguishable particle ensembles,'' {\em arXiv preprint
  arXiv:1601.02703}, 2016.

\bibitem{khachaturyan2013theory}
A.~G. Khachaturyan, {\em Theory of structural transformations in solids}.
\newblock Courier Corporation, 2013.

\bibitem{girifalco2003statistical}
L.~A. Girifalco, {\em Statistical mechanics of solids}, vol.~58.
\newblock OUP USA, 2003.

\bibitem{weisstein2008stirling}
E.~W. Weisstein, ``Stirling's approximation,'' {\em MathWorld.
  http://mathworld.wolfram.com/StirlingsApproximation.html}, 2008.

\bibitem{lesar2013introduction}
R.~LeSar, {\em Introduction to computational materials science: fundamentals to
  applications}.
\newblock Cambridge University Press, 2013.

\bibitem{beretta2017steepest}
G.~P. Beretta, O.~Al-Abbasi, and M.~R. von Spakovsky, ``Steepest-entropy-ascent
  nonequilibrium quantum thermodynamic framework to model chemical reaction
  rates at an atomistic level,'' {\em Physical Review E}, vol.~95, no.~4,
  p.~042139, 2017.

\bibitem{li2018multiscale}
G.~Li, M.~R. von Spakovsky, F.~Shen, and K.~Lu, ``Multiscale {T}ransient and
  {S}teady-{S}tate {S}tudy of the {I}nfluence of {M}icrostructure {D}egradation
  and {C}hromium {O}xide {P}oisoning on {S}olid {O}xide {F}uel {C}ell {C}athode
  {P}erformance,'' {\em Journal of Non-Equilibrium Thermodynamics}, vol.~43,
  no.~1, pp.~21--42, 2018.

\bibitem{beretta2014steepest}
G.~P. Beretta, ``{S}teepest entropy ascent model for far-nonequilibrium
  thermodynamics: {U}nified implementation of the maximum entropy production
  principle,'' {\em Physical Review E}, vol.~90, no.~4, p.~042113, 2014.

\bibitem{li2018steepest}
G.~Li, M.~R. von Spakovsky, and C.~Hin, ``Steepest entropy ascent quantum
  thermodynamic model of electron and phonon transport,'' {\em Physical Review
  B}, vol.~97, no.~2, p.~024308, 2018.

\bibitem{nishizawa1984co}
T.~Nishizawa and K.~Ishida, ``The {C}o-{C}u ({C}obalt-{C}opper) {S}ystem,''
  {\em Bulletin of Alloy Phase Diagrams}, vol.~5, no.~2, pp.~161--165, 1984.

\bibitem{legoues1984influence}
F.~K. LeGoues and H.~I. Aaronson, ``Influence of crystallography upon critical
  nucleus shapes and kinetics of homogeneous f.c.c-f.c.c nucleation--{IV}.
  {C}omparisons between theory and experiment in {C}u--{C}o alloys,'' {\em Acta
  Metallurgica}, vol.~32, no.~10, pp.~1855--1864, 1984.

\bibitem{wendt1985atom}
H.~Wendt and P.~Haasen, ``Atom probe field ion microscopy of the decomposition
  of {C}u-2.7 at\% {C}o,'' {\em Scripta metallurgica}, vol.~19, no.~9,
  pp.~1053--1058, 1985.

\bibitem{hattenhauer1993decomposition}
R.~Hattenhauer and P.~Haasen, ``The decomposition kinetics of {C}u--1 at.\%
  {C}o at 823\;{K}, studied by bright-field-zone-axis-incidence transmission
  electron microscopy,'' {\em Philosophical Magazine A}, vol.~68, no.~6,
  pp.~1195--1213, 1993.

\bibitem{busch1996high}
R.~Busch, F.~G{\"a}rtner, C.~Borchers, P.~Haasen, and R.~Bormann, ``High
  resolution microstructure analysis of the decomposition of
  {C}u$_{90}${C}o$_{10}$ alloys,'' {\em Acta materialia}, vol.~44, no.~6,
  pp.~2567--2579, 1996.

\bibitem{dohl1984measurement}
R.~D{\"o}hl, M.-P. Macht, and V.~Naundorf, ``Measurement of the diffusion
  coefficient of cobalt in copper,'' {\em physica status solidi (a)}, vol.~86,
  no.~2, pp.~603--612, 1984.

\bibitem{liu1993spinodal}
J.-M. Liu, R.~Busch, F.~G{\"a}rtner, P.~Haasen, Z.-G. Liu, and Z.-C. Wu,
  ``Spinodal decomposition of {C}u{C}o alloys,'' {\em physica status solidi
  (a)}, vol.~138, no.~1, pp.~157--174, 1993.

\end{thebibliography}

\end{document}